\documentclass{aip-cp}

\usepackage[numbers]{natbib}
\usepackage{rotating}
\usepackage{graphicx}

\begin{document}

\title{Recent Results on Hadron Spectroscopy from BESIII}

\author[aff1]{Beijiang Liu\corref{cor1} (for the BESIII collaboration)}

\affil[aff1]{Institute of High Energy Physics, Chinese Academy of Sciences, 19B Yuanquanlu, Shijingshan district, Beijing, 100049,
China}
\corresp[cor1]{Corresponding author: liubj@ihep.ac.cn}

\maketitle

\begin{abstract}
Hadron spectroscopy is one of the most important physics goals of BESIII. BESIII brings great opportunities to study the XYZ states of charmonium by directly producing the Y states up to 4.6~GeV. High statistics of charmonium decays collected at BESIII provide an excellent place for hunting gluonic excitations and studying the excited baryons. Recent results of light hadron spectroscopy and charmonium spectroscopy from BESIII will be reported.
\end{abstract}

\section{INTRODUCTION}
Hadron spectroscopy is a unique way to access Quantum Chromodynamics (QCD). QCD-motivated models for hadrons predict an assortment of ``Exotic Hadrons'' that have structures that are more complex than the quark-antiquark mesons and three-quark baryons of the original quark model, such as glueballs, hybrids and multi-quark states. Experimental search of these predictions and subsequent investigation of their properties would provide validation of and valuable input to the quantitative understanding of QCD.

BESIII (Beijing Spectrometer) is a general purpose $4\pi$ detector at the
upgraded BEPCII (Beijing Electron and
Positron Collider) that operated in the $\tau$-charm threshold energy region\cite{bes3dect}.
Since 2009, it has collected the world's
largest data samples of $J/\psi$, $\psi(3686)$,
$\psi(3770)$ and $\psi(4040)$ decays. More recently, data were taken in the energy region above 4 GeV,
where energies up to about 4.6 GeV are accessible. These data
are being used to make a variety of
interesting and unique studies of light
hadron spectroscopy, charmonium spectroscopy, high-statistics
measurements of charmonium decays and D meson decays.

\section{Charmonium spectroscopy}
The quark model~\cite{Fritzsch:1973pi}, which treats mesons as combinations of one quark and one anti-quark,
is very successful in describing meson properties, particularly in the charmonium ($c \bar{c}$)
region below the open-charm threshold.~\cite{Barnes:2005pb}. The past decade, however, has seen the discovery
of a number of new states (named the ¡°XYZ¡± states) that do not fit within this model, and
which perhaps point towards more complex systems. With its unique data samples at energies of 3.8--4.6~GeV, the
BESIII experiment made a significant contribution to the study of
charmonium and charmonium-like states.

\subsection{Charged charmonum-like states: $Z_c$'s}
\begin{figure}[htb]
\centering
     \includegraphics[width=0.32\textwidth,height=0.19\textheight]{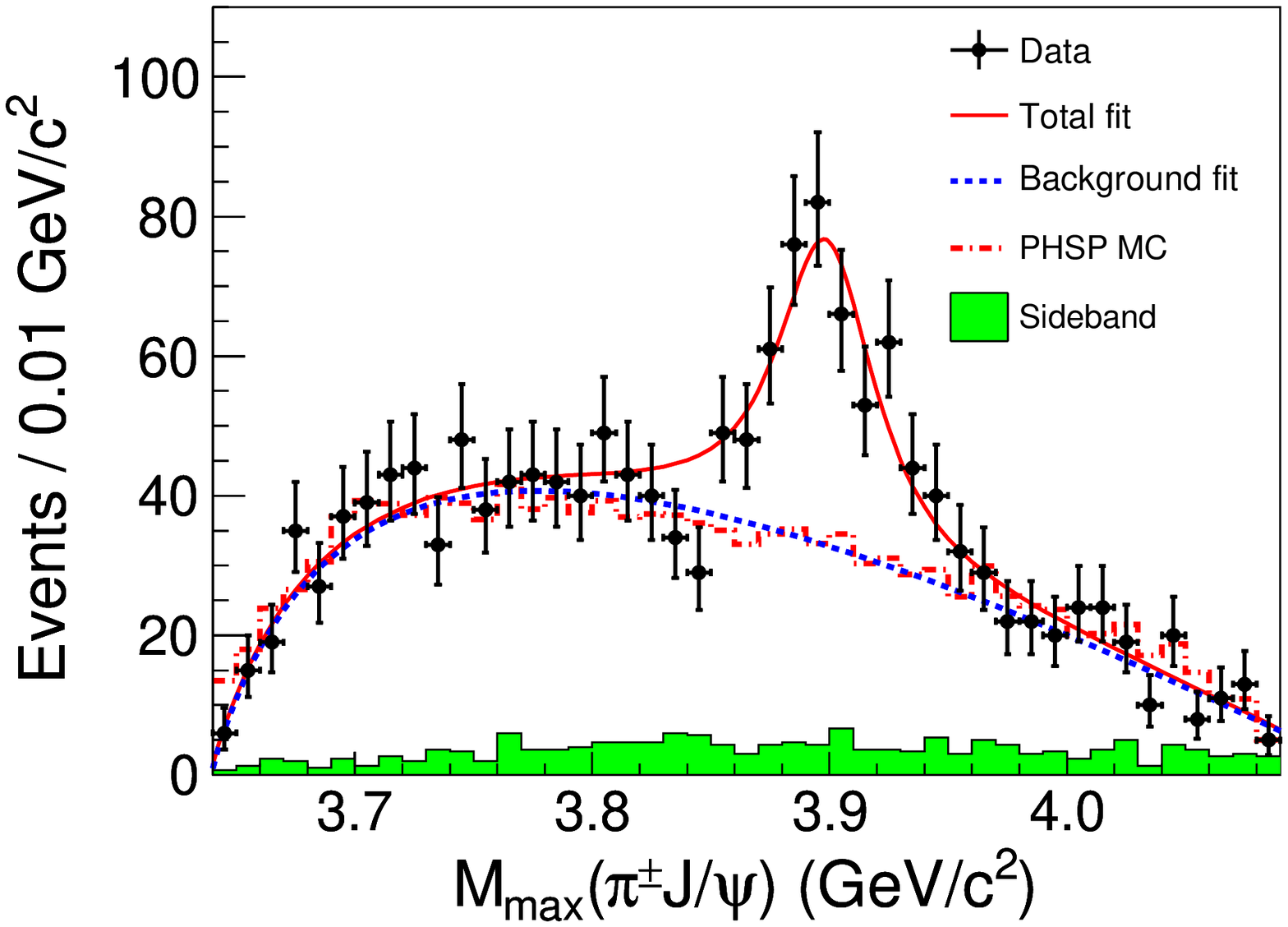}
     \put(-135,3){(a)}
     \includegraphics[width=0.32\textwidth,height=0.19\textheight]{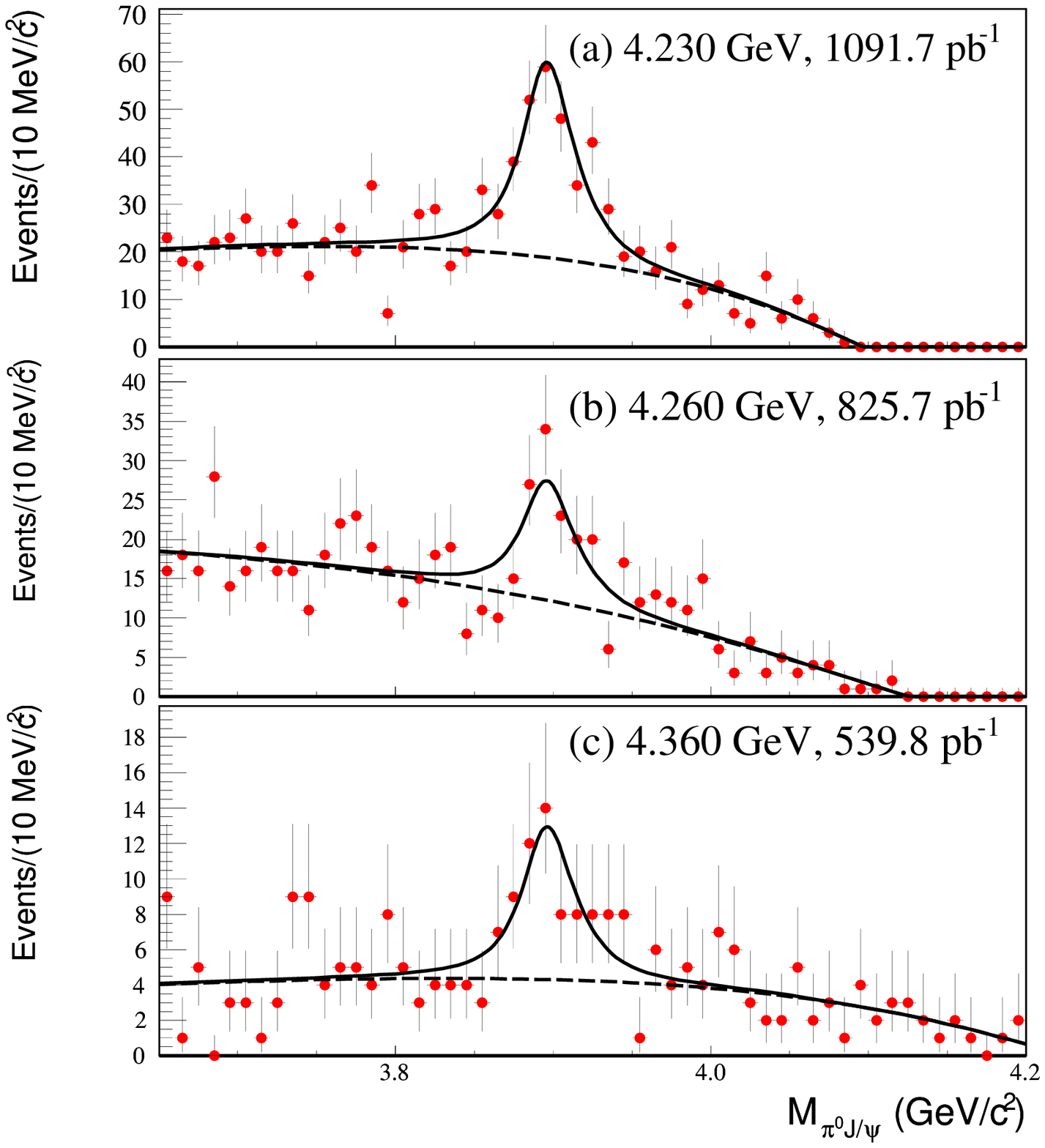}
     \put(-135,3){(b)}
\caption{$Z_c(3900)\to \pi J/\psi$ production in $e^+e^- \to \pi \pi J/\psi$ processes:
(a) $Z_c(3900)^{\pm} \to \pi^{\pm} J/\psi$~\cite{Ablikim:2013mio}, (b) $Z_c(3900)^0 \to \pi^{0} J/\psi$~\cite{BESIII:2015kha}.
}
\label{fig:Zc3900pmz}
\end{figure}

\begin{figure}[htb]
\centering
     \includegraphics[width=0.32\textwidth,height=0.19\textheight]{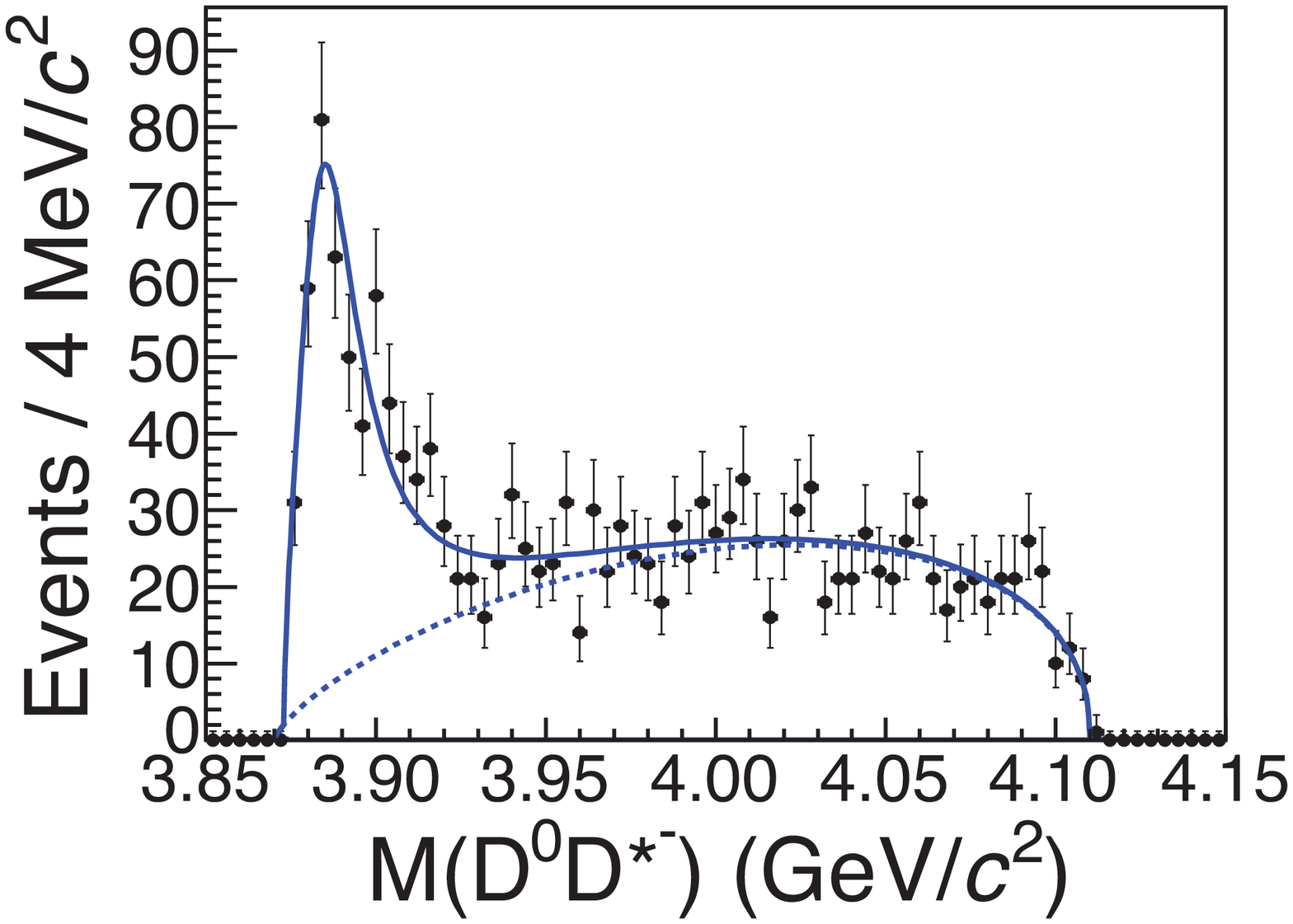}
     \put(-135,3){(a)}
     \includegraphics[width=0.32\textwidth,height=0.19\textheight]{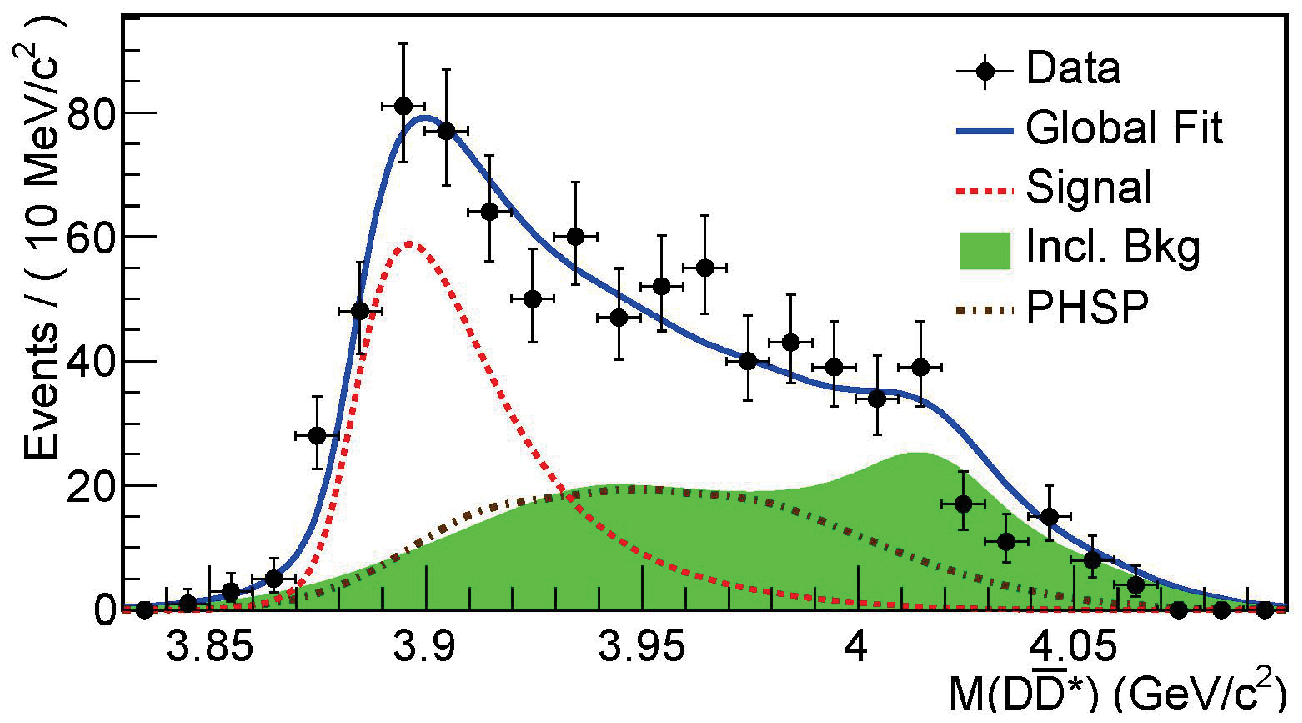}
     \put(-135,3){(b)}
\caption{$Z_c(3885)  \to (D \overline{D}^*) $  production in $e^+e^- \to \pi (D \overline{D}^*) $ processes:
(a) $Z_c(3885)^{\pm} \to (D \overline{D}^*)^{\pm}$ ~\cite{Ablikim:2013xfr}. (b) $Z_{c}(3885)^{0}\to
  D\bar{D}^{*}$~\cite{Ablikim:2015gda}.
}
\label{fig:Zc3885pmz}
\end{figure}

Recently, in the study of $e^+e^-\to J/\psi \pi^+\pi^-$, a distinct
charged structure, named the $Z_c(3900)^{\pm}$, was observed in the
$J/\psi\pi^\pm$ spectrum by BESIII~\cite{Ablikim:2013mio} and
Belle~\cite{Liu:2013dau}.  Its existence was confirmed shortly thereafter with CLEO-c
data~\cite{Xiao:2013iha}. The existence of the neutral partner in the
decay $Z_c(3900)^0\to J/\psi\pi^0$ has also been reported in CLEO-c
data~\cite{Xiao:2013iha} and by BESIII~\cite{BESIII:2015kha},
thus complementing the isospin-triplet representation of isospin one, $I=1$, resonances. The
$Z_c(3900)$ is a good candidate for an exotic state beyond simple quark models,
since it contains
a $c\bar{c}$ pair and is also electrically charged. Noting that the
$Z_c(3900)$ has a mass very close to the $D^*\bar{D}$ threshold
(3875~MeV), BESIII analyzed the process
$e^{+}e^{-}\to\pi^{\pm}(D\bar{D}^{*})^{\mp}$, and a clear structure in
the $(D\bar{D}^{*})^{\mp}$ mass spectrum is seen, called the $Z_c(3885)$. The
measured mass and width are $(3883.9 \pm 1.5 \pm 4.2)$~MeV/$c^2$ and
$(24.8\pm 3.3 \pm 11.0)$~MeV, respectively, and quantum numbers $J^P =
1^+$ are favored~\cite{Ablikim:2013xfr}. A neutral structure in the $D\bar{D}^{*}$ system around the $D\bar{D}^{*}$ mass threshold is observed
  with a statistical significance greater than 10$\sigma$ in the
  processes $e^{+}e^{-}\rightarrow D^{+}D^{*-}\pi^{0}+c.c.$ and
  $e^{+}e^{-}\rightarrow D^{0}\bar{D}^{*0}\pi^{0}+c.c.$ at $\sqrt{s}$
  = 4.226 and 4.257 GeV in the BESIII experiment~\cite{Ablikim:2015gda}. Assuming the $Z_c(3885)\to
D\bar{D}^{*}$ and the $Z_c(3900)\to J/\psi\pi$ signals are from the
same source, the ratio of partial widths $\frac{\Gamma(Z_c(3885)\to
  D\bar{D}^{*})}{\Gamma(Z_c(3900)\to J/\psi\pi)}$ is determined to be $6.2\pm1.1\pm2.7$. This ratio
is much smaller than typical values for decays of conventional
charmonium states above the open charm threshold. BESIII also searched for the decay mode of $Z_c$ with the annihilation of
$c\bar{c}$ in the reaction
$e^+e^- \to \omega \pi^+ \pi^-$~\cite{Ablikim:2015cag}. The upper limits of the Born cross section are determined to be 0.26 and 0.18 pb, at $E_{CM}$ = 4.23 and 4.26 GeV, respectively. Figure~\ref{fig:Zc3900pmz} shows the $Z_c(3900)^{\pm} \to \pi^{\pm} J/\psi$ and its isospin-partner,  $Z_c(3900)^0 \to \pi^{0} J/\psi$. The results of $Z_c(3885)  \to (D \overline{D}^*) $  observed in $e^+e^- \to \pi (D \overline{D}^*) $ processes are shown in Fig.~\ref{fig:Zc3885pmz}.Table~\ref{tab:Zc3900_mw} lists corresponding masses and widths of the $Z_c$ states near the $D^*\bar{D}$ threshold.

\begin{figure}[htb]
\centering
     \includegraphics[width=0.32\textwidth,height=0.19\textheight]{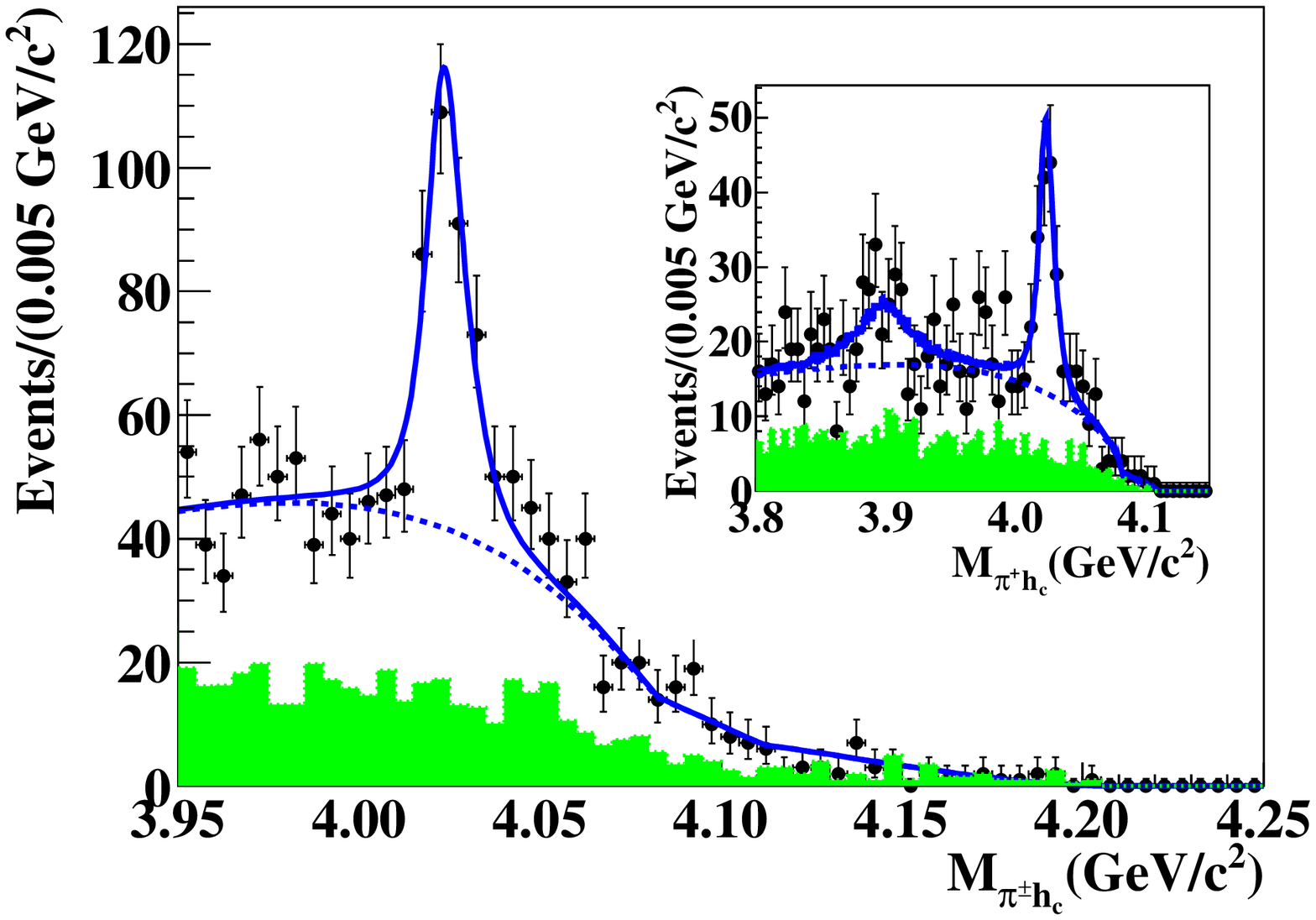}
     \put(-135,3){(a)}
     \includegraphics[width=0.32\textwidth,height=0.19\textheight]{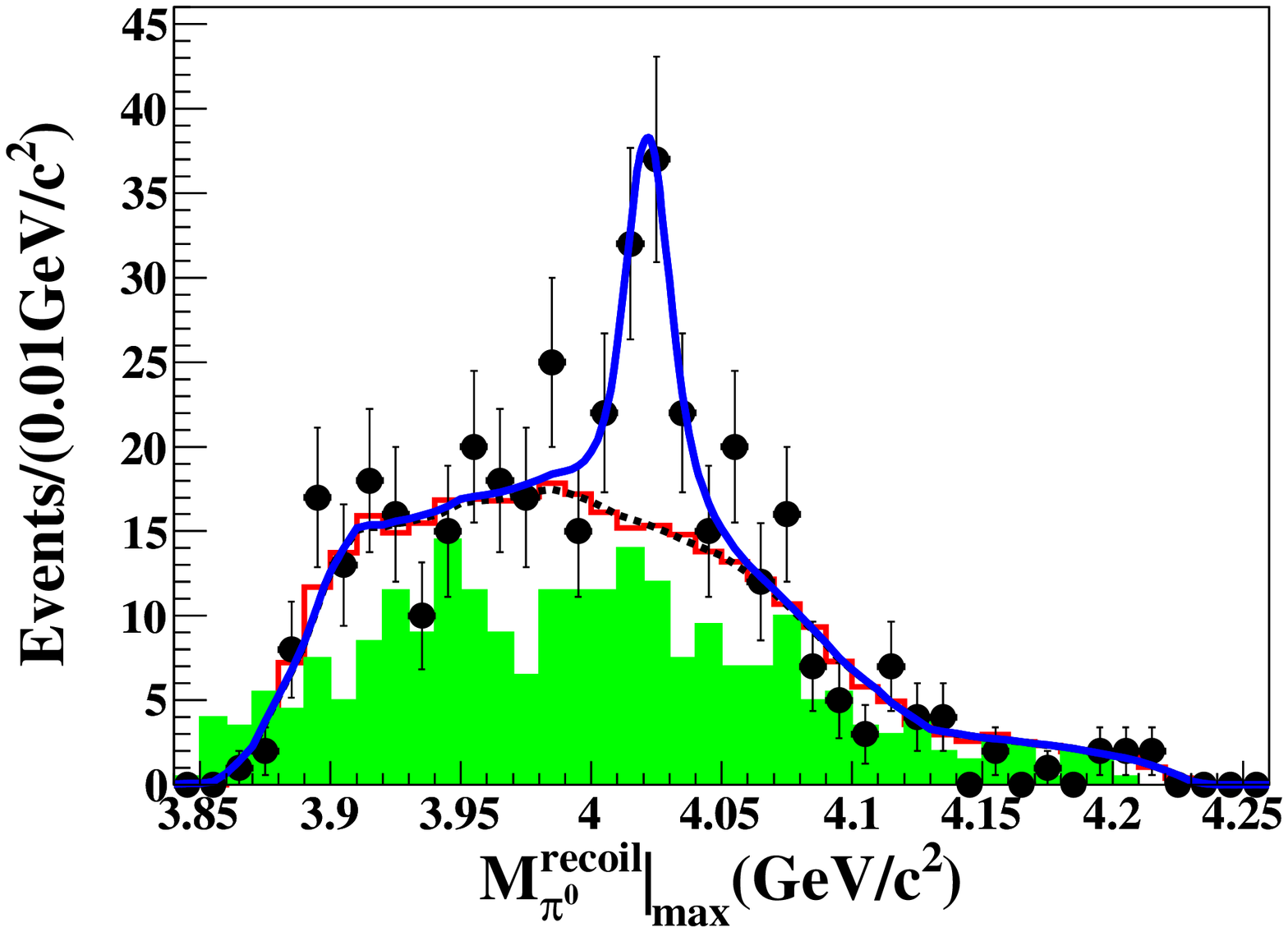}
     \put(-135,3){(b)}
\caption{$Z_c(4020)  \to \pi  h_c$ production in $e^+e^- \to \pi \pi h_c$ processes: (a)charged mode~\cite{Ablikim:2013wzq}, (b) neutral mode~\cite{Ablikim:2014dxl}.
}
\label{fig:Zc4020pmz}
\end{figure}

\begin{figure}[htb]
\centering
      \includegraphics[width=0.32\textwidth,height=0.19\textheight]{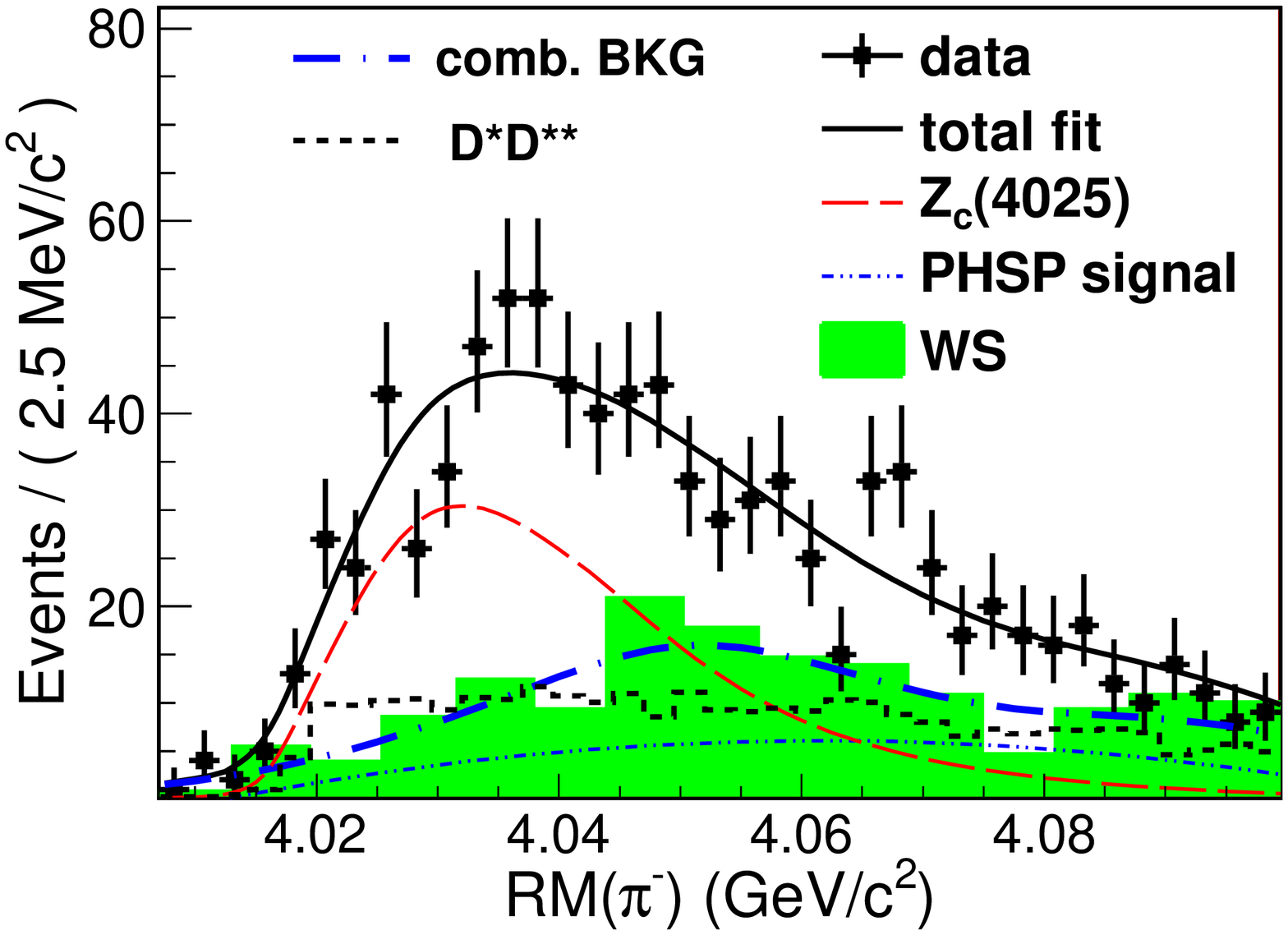}
     \put(-135,3){(a)}
     \includegraphics[width=0.32\textwidth,height=0.19\textheight]{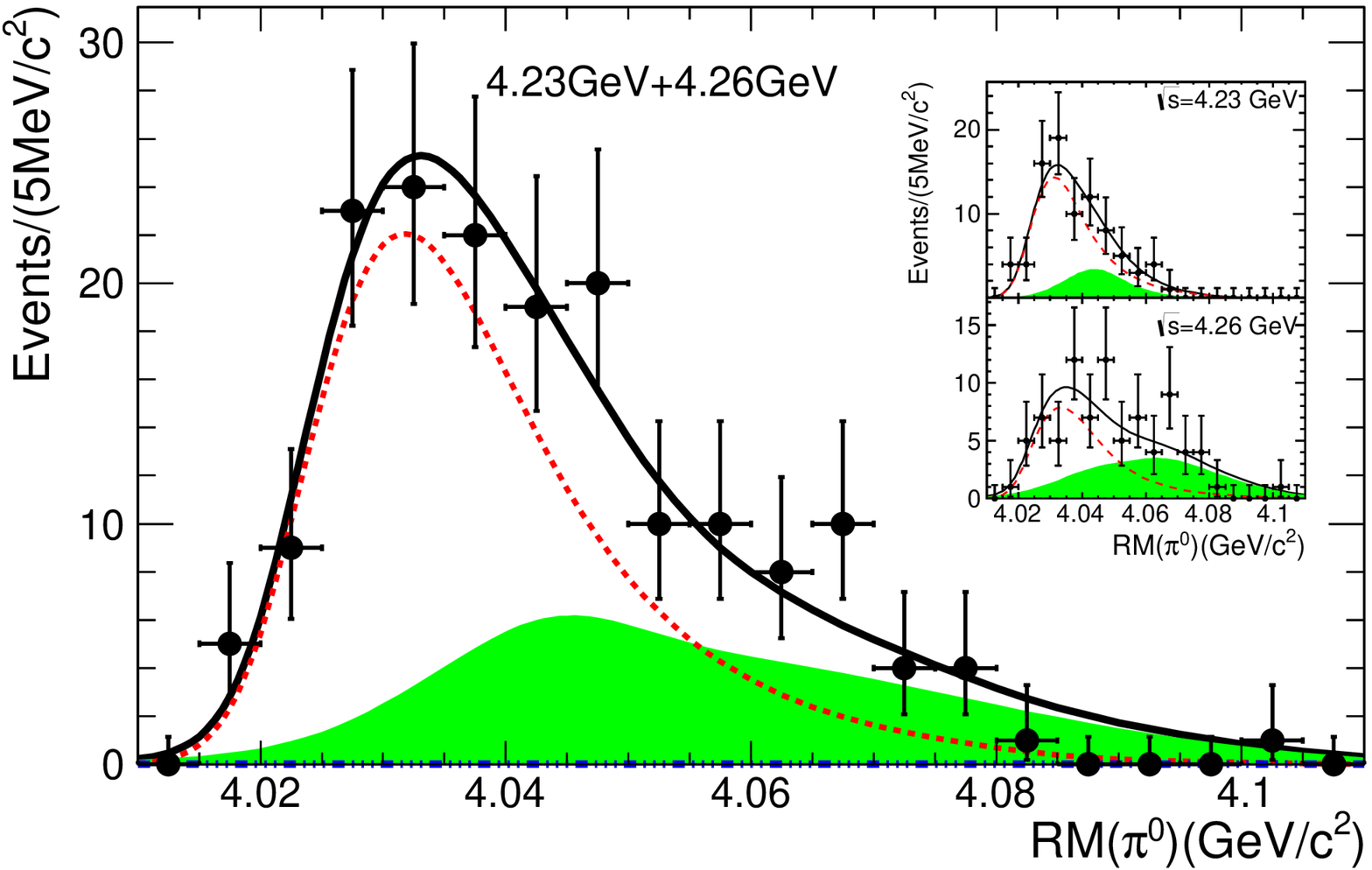}
     \put(-135,3){(b)}
\caption{ $Z_c(4025)  \to (D^* \overline{D} ^*)$ production in $e^+e^- \to \pi (D^* \overline{D} ^*)$ processes: (a) $Z_c(4025)^{\pm} \to (D^* \overline{D} ^*)^{\pm}$ ~\cite{Ablikim:2013emm}, (b) $Z_c(4025)^{0} \to (D^* \overline{D} ^*)^{0}$ ~\cite{Ablikim:2015vvn}.
}
\label{fig:Zc4025pmz}
\end{figure}

BESIII analyzed the  reaction
$e^+e^- \to \pi^+ \pi^- h_c$, at $E_{CM}$ = 4.23, 4.26 and 4.36~GeV~\cite{Ablikim:2013wzq}.
Corresponding $\pi^{\pm} h_c$ invariant mass distribution, when all energy points are combined, is shown in Fig.~\ref{fig:Zc4020pmz}.
The $Z_c(4020)^{\pm}$ signal has 8.9$\sigma$ significance. And the inset represents a search for the $Z_c(3900) \to \pi h_c$, at $E_{CM}$ = 4.23 and 4.26~GeV. At the 90\% confidence level (C.L.), the
upper limits on the production cross-sections are set to
$\sigma(e^+e^-\to \pi^\pm Z_c(4020)^\mp\to \pi^+ \pi^- h_c) <13$~pb at 4.23~GeV and
$<11$~pb at 4.26~GeV. These are lower than those of $Z_c(4020)\to
\pi^\pm J/\psi$~\cite{Ablikim:2013mio}.
BESIII also observed $e^+e^- \to \pi^0 \pi^0 h_c$ at $\sqrt{s}=4.23$,
4.26, and 4.36~GeV~\cite{Ablikim:2014dxl}, confirming the isospin of $Z_c(4020)$ to
be one.
Similarly to the case of the $Z_c(3900)$, the mass of the $Z_c(4020)$ is near
the threshold for $D^{*} \overline{D}^{*} $ production.
BESIII studied two reactions: $e^+e^- \to (D^{*} \overline{D} ^{*})^{\pm} \pi^{\mp}$,
at $E_{CM}$ = 4.26~GeV~\cite{Ablikim:2013emm},
and $e^+e^- \to (D^{*} \overline{D} ^{*})^{0} \pi^{0}$,
at $E_{CM}$ = 4.23 and 4.26~GeV~\cite{Ablikim:2015vvn}. A structure that couples to $D^{*} \overline{D^{*}}$
is evident, in both charged and neutral decays, denoted as $Z_c(4025)$. Fig.~\ref{fig:Zc4025pmz} shows the results of $Z_c(4025)$. Assuming the $Z_c(4020)$ and the $Z_c(4025)$ signals are from the
same source, the ratio of partial widths $\frac{\Gamma(Z_c(4025)\to
  D^*\bar{D}^{*})}{\Gamma(Z_c(4020)\to h_c\pi)}$ is determined to be $12\pm5$. Table~\ref{tab:Zc3900_mw} lists corresponding masses and widths of the $Z_c$ states near the $D^{*} \overline{D^{*}}$ threshold.

\begin{table}[htb]
\begin{tabular}{l|cc|c}
Decay &  Mass [MeV/$c^2$] &  Width [MeV]  & Ref.  \\
\hline
$Z_c(3900)^{\pm} \to \pi^{\pm} J/\psi$ &
 3899.0$\pm$3.6$\pm$4.9 & 46$\pm$10$\pm$20 &  \cite{Ablikim:2013mio} \\
$Z_c(3900)^{0} \to \pi^{0} J/\psi$ &
3894.8$\pm$2.3$\pm$3.2 & 29.6$\pm$8.2$\pm$8.2 & \cite{BESIII:2015kha} \\

\hline
$Z_c(3885)^{\pm} \to (D \overline{D}^*)^{\pm}$ (single D-tag) &
  3883.9$\pm$1.5$\pm$4.2 & 24.8$\pm$3.3$\pm$11.0 &  ~\cite{Ablikim:2013xfr} \\
$Z_c(3885)^{\pm} \to (D \overline{D}^*)^{\pm}$ (double D-tag) &
  3881.7$\pm$1.6$\pm$2.1 & 26.6$\pm$2.0$\pm$2.3 & ~\cite{Ablikim:2015swa} \\
$Z_c(3885)^{0} \to (D \overline{D}^*)^{0}$ &
  3885.7$^{4.3}_{-5.7}\pm$9.4 & 35$^{+11}_{-12}\pm$15 & ~\cite{Ablikim:2015gda} \\

\hline
$Z_c(4020)^{\pm} \to \pi^{\pm} h_c$ &
4022.9$\pm$0.8$\pm$2.7 & 7.9$\pm$2.7$\pm$2.6 & \cite{Ablikim:2013wzq} \\
$Z_c(4020)^{0} \to \pi^{0} h_c$ &
  4023.9$\pm$2.2$\pm$3.8 & fixed &  \cite{Ablikim:2014dxl} \\
\hline

$Z_c(4025)^{\pm} \to (D^* \overline{D} ^*)^{\pm}$ &
4026.3$\pm$2.6$\pm$3.7 & 24.8$\pm$5.6$\pm$7.7 &  \cite{Ablikim:2013emm} \\
$Z_c(4025)^{0} \to (D^* \overline{D} ^*)^{0}$ &
4025.5$^{+2.0}_{-4.7}\pm3.1$ & 23.0$\pm$6.0$\pm$1.0 & \cite{Ablikim:2015vvn} \\

\hline
\end{tabular}
\caption{Masses and widths of exotic meson candidates studied by BESIII.}
\label{tab:Zc3900_mw}
\end{table}

\subsection{Observation of $Y(4260)\to\gamma X(3872)$}

\begin{figure}[htb]
   \centering
     \includegraphics[width=0.32\textwidth,height=0.19\textheight]{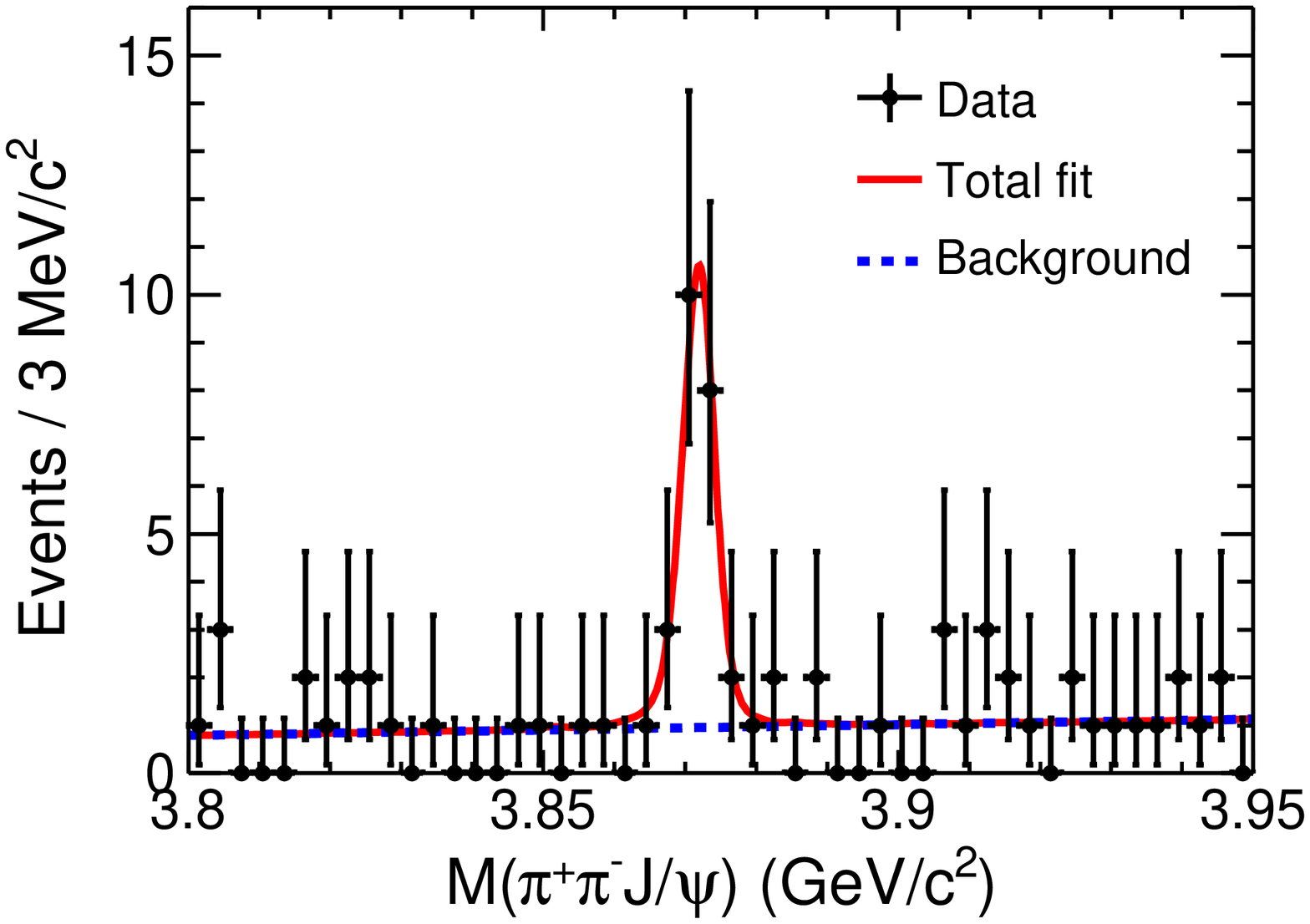}
     \put(-135,3){(a)}
     \includegraphics[width=0.32\textwidth,height=0.19\textheight]{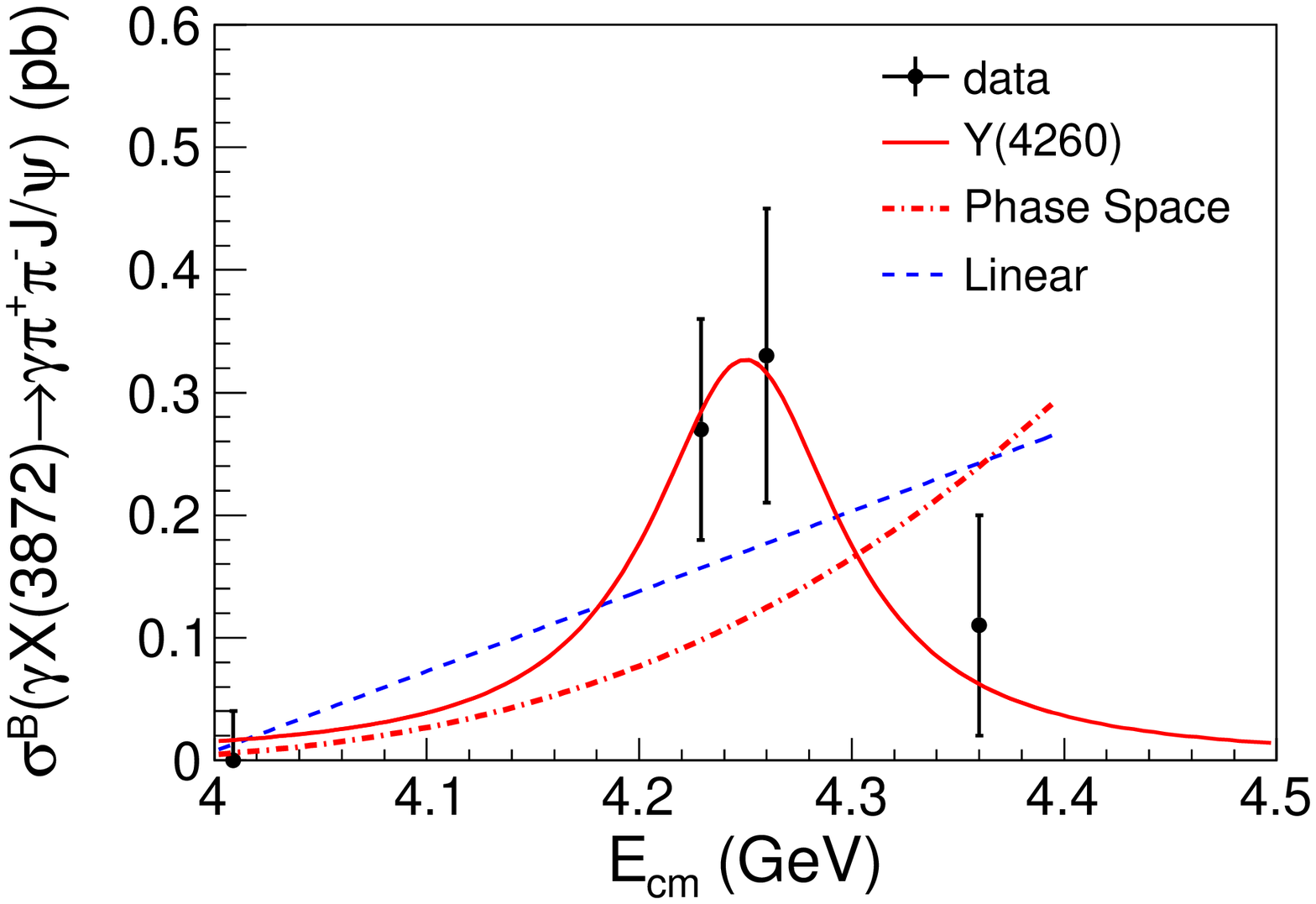}
     \put(-135,3){(b)}
\caption{(a) Fit to the $M(\pi^+\pi^- J/\psi)$ distribution observed at
BESIII. Dots with error bars are data, the curves are the best
fit. (b) Fit to $\sigma^B[e^+e^-\to \gamma X(3872)]\times
\mathcal{B}[X(3872)\to \pi^+\pi^- J/\psi]$ measured by BESIII with a $Y(4260)$
resonance (red solid curve), a linear continuum (blue dashed
curve), or an $E1$-transition phase space term (red dotted-dashed
curve). Dots with error bars are data.} \label{fit-mx}
\end{figure}

BESIII observed $e^+e^-\to \gamma X(3872)\to \gamma \pi^+\pi^- J/\psi$, with $J/\psi$
reconstructed through its decays into lepton pairs ($\ell^+\ell^-=e^+e^-$ or
$\mu^+\mu^-$)~\cite{Ablikim:2013dyn}.The $M(\pi^+\pi^- J/\psi)$ distribution (summed over all energy points), as
shown in Fig.~\ref{fit-mx}~(left), was fitted to extract the mass
and signal yield of $X(3872)$. The Born-order cross-section was measured. The energy-dependent cross-sections were fitted with a $Y(4260)$
resonance (parameters fixed to PDG~\cite{pdg2014} values), linear
continuum, or $E1$-transition phase space ($\propto E^3_\gamma$)
term, as shown in Figure~\ref{fit-mx}~(right). The
measurements are consistent with expectations for the radiative transition
process $Y(4260)\to \gamma X(3872)$. Combining the above with the $e^+e^-\to \pi^+\pi^- J/\psi$ cross-section
measurement at $\sqrt{s}=4.26$~GeV from BESIII~\cite{Ablikim:2013mio}, we
obtain $\sigma^B[e^+e^-\to \gamma X(3872)]\cdot {\cal B}[X(3872)\to
\pi^+\pi^- J/\psi]/\sigma^B(e^+e^-\to \pi^+\pi^- J/\psi) = (5.2\pm 1.9)\times 10^{-3}$,
under the assumption that $X(3872)$ and $\pi^+\pi^- J/\psi$ are only produced
from $Y(4260)$ decays. If we take ${\cal B}[X(3872)\to \pi^+\pi^- J/\psi] =
5\%$~\cite{bnote}, then $\mathcal{R} = \frac{{\cal B}[Y(4260)\to
\gamma X(3872)]}{{\cal B}(Y(4260)\to \pi^+\pi^- J/\psi)}\sim 0.1$.

\subsection{Observation of $\psi(1\,^3D_2)$}

\begin{figure}[htb]
\centering
     \includegraphics[width=0.32\textwidth,height=0.19\textheight]{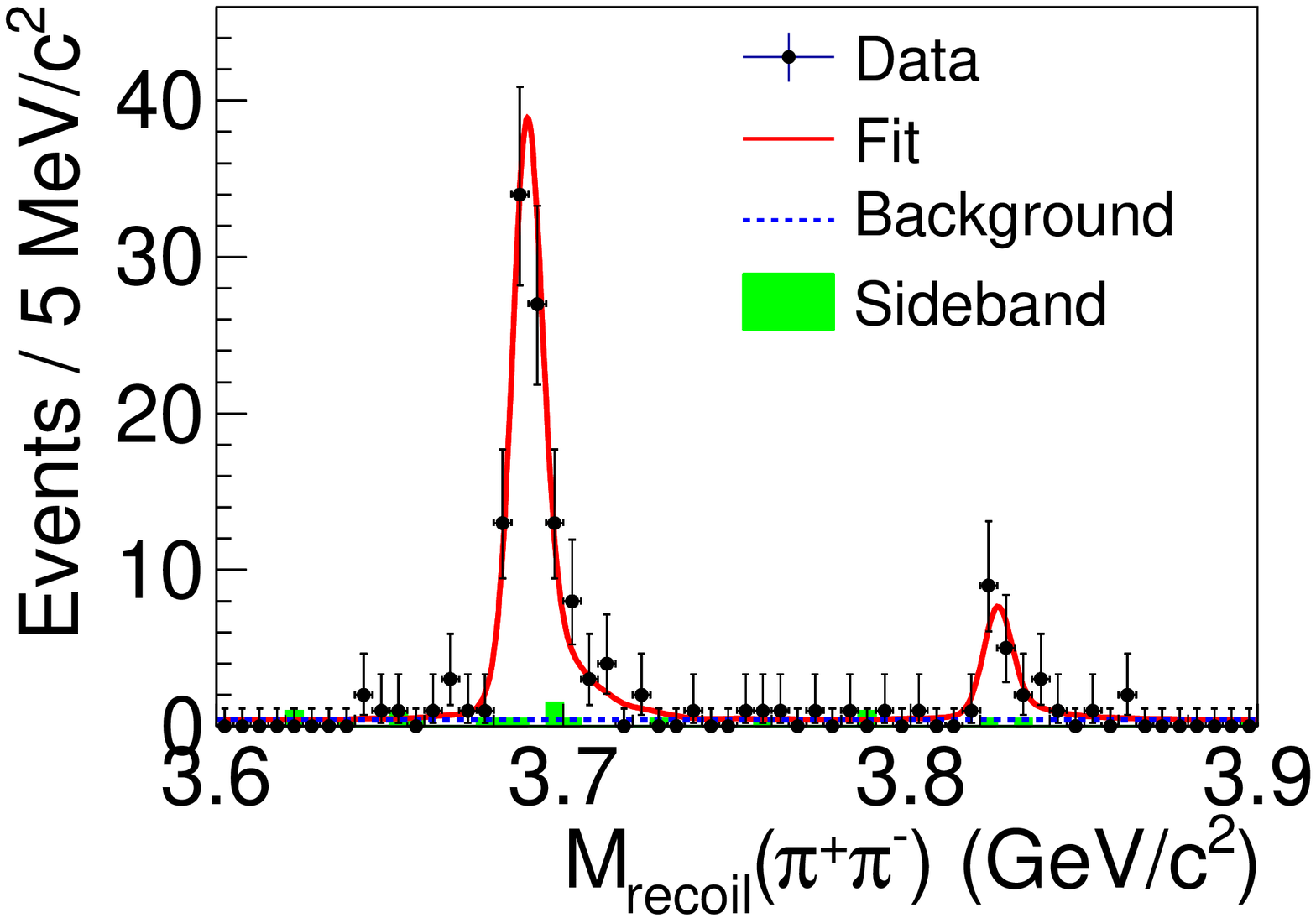}
     \put(-135,6){(a)}
     \includegraphics[width=0.32\textwidth,height=0.19\textheight]{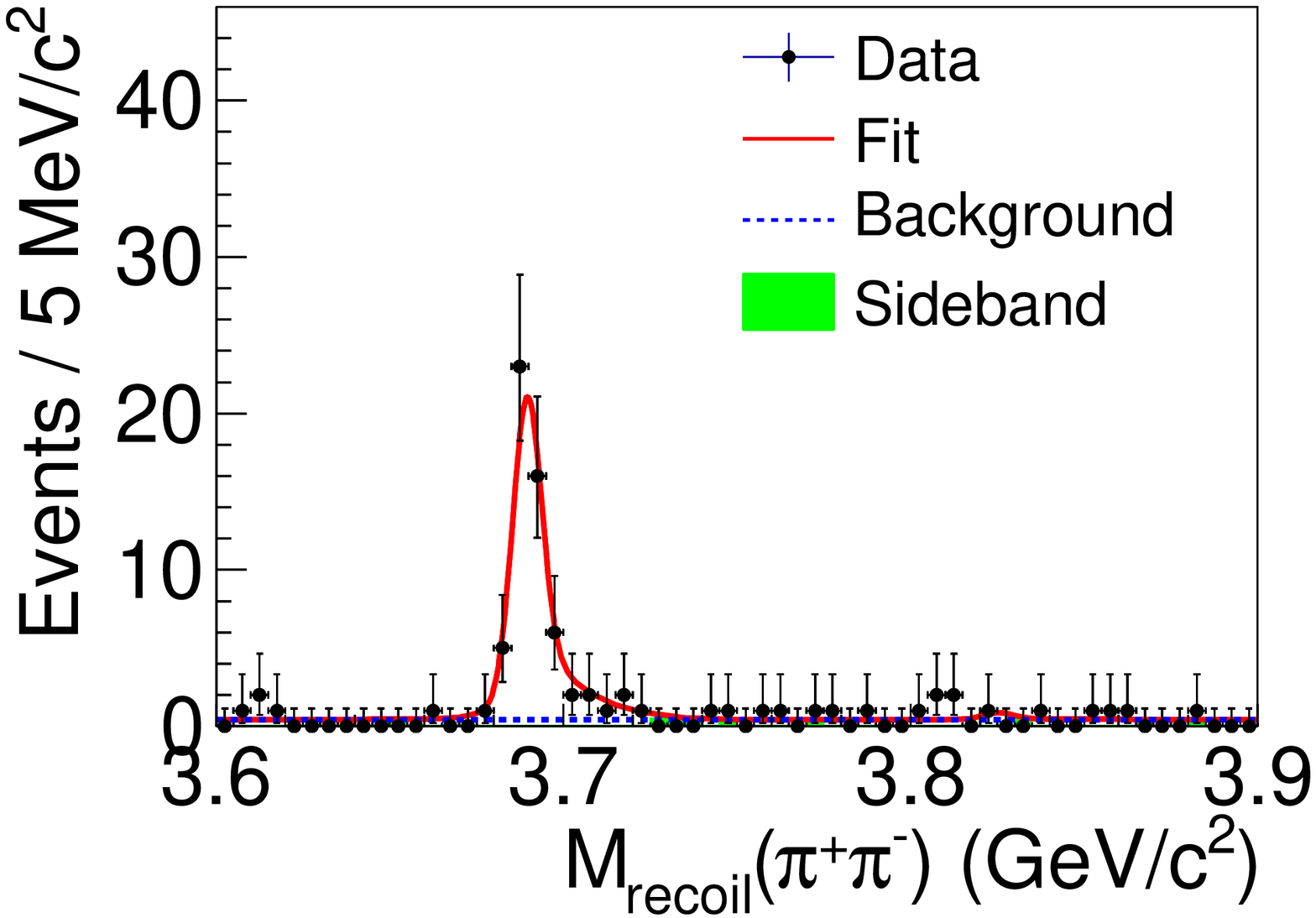}
     \put(-135,6){(b)}
     \includegraphics[width=0.32\textwidth,height=0.19\textheight]{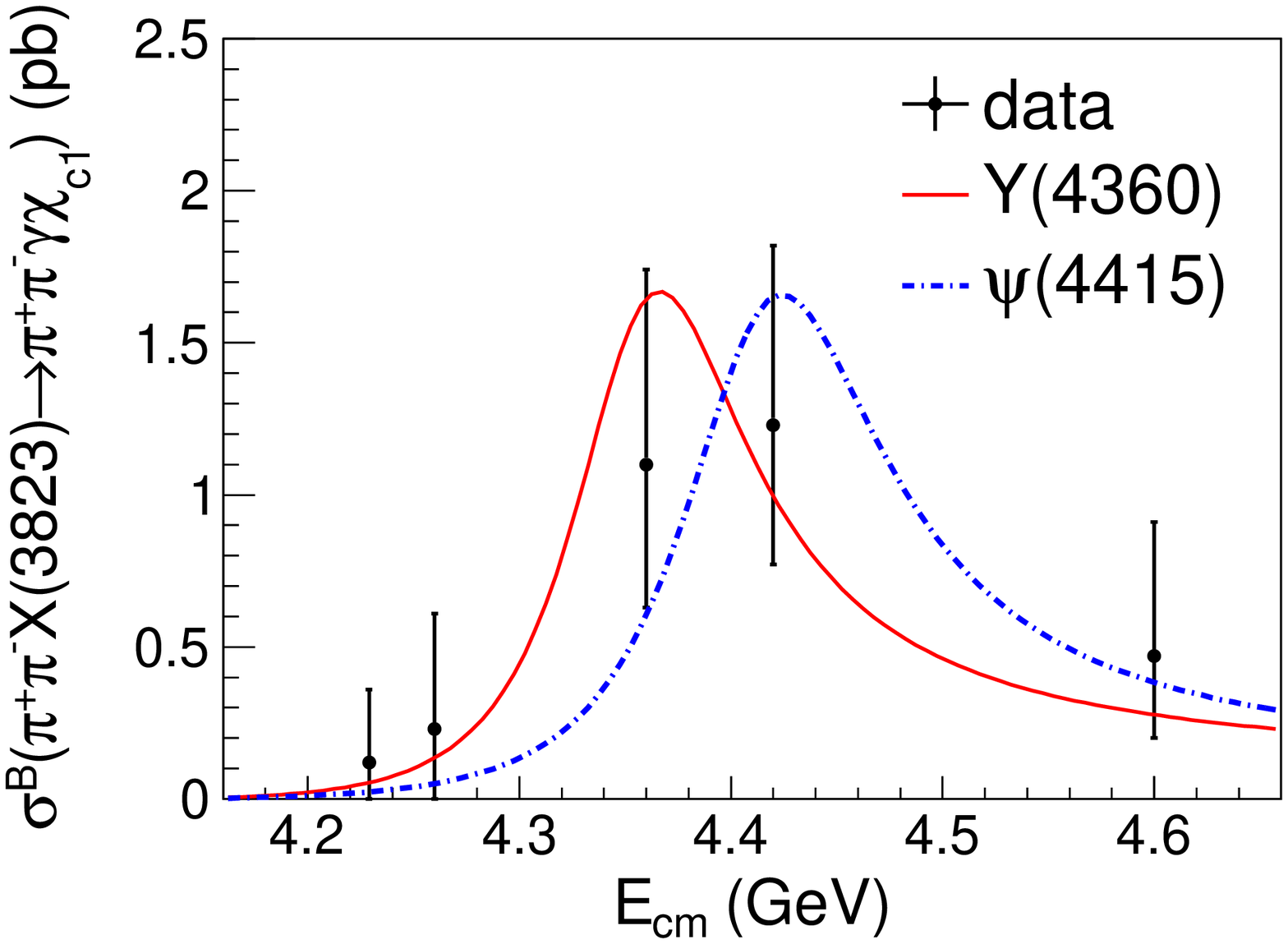}
     \put(-135,6){(c)}
\caption{Simultaneous fit to the $M_{\rm recoil}(\pi^+\pi^-)$
distribution of $\gamma\chi_{c1}$ events (a) and $\gamma\chi_{c2}$ events (b). Dots with error
bars are data, red solid curves are total fit, dashed blue curves
are background, and the green shaded histograms are $J/\psi$ mass
sideband events. (c) Comparison of the energy-dependent cross sections of
$\sigma^B[e^+e^-\to\pi^+\pi^- X(3823)]\cdot
\mathcal{B}(X(3823)\to\gamma\chi_{c1})$ to the $Y(4360)$ and
$\psi(4415)$ line shapes. Dots with error bars (statistical only)
are data.  The red solid (blue dashed) curve shows a fit with the
$Y(4360)$ ($\psi(4415)$) line shape.} \label{X-fit}
\end{figure}

BESIII observed $X(3823)$ in the $e^+e^-\to \pi^+\pi^-X(3823) \to
\pi^+\pi^-\gamma\chi_{c1}$ process with a statistical significance
of $6.2\sigma$ in data samples at c.m. energies of
$\sqrt{s}=$4.23, 4.26, 4.36, 4.42, and 4.60~GeV~\cite{Ablikim:2015dlj}. The measured mass of the $X(3823)$ is
$(3821.7\pm 1.3\pm 0.7)$~MeV/$c^2$, and the width is less than
$16$~MeV at the 90\% confidence level. The products
of the Born cross sections for $e^+e^-\to \pi^+\pi^-X(3823)$
and the branching ratio $\mathcal{B}[X(3823)\to \gamma\chi_{c1,c2}]$ are
also measured. These measurements are in good agreement with
the assignment of the $X(3823)$ as the $\psi(1^3D_2)$ charmonium
state. The fitted results to $\pi^+\pi^-$ recoil mass
distributions for events in the $\chi_{c1}$ and $\chi_{c2}$ signal
regions are shown in Fig.~\ref{X-fit} (a) and Fig.~\ref{X-fit} (b), respectively.

The production cross-sections of $\sigma^{B}(e^+e^-\to\pi^+\pi^-
X(3823))\cdot \mathcal{B}(X(3823)\to \gamma\chi_{c1}$,
$\gamma\chi_{c2})$ were also measured at these c.m. energies. The
cross-sections of $e^+e^-\to\pi^+\pi^- X(3823)$ were fitted with the
$Y(4360)$ shape or the $\psi(4415)$ shape, as shown in Fig.~\ref{X-fit} (c). Both the
$Y(4360)$ and $\psi(4415)$ hypotheses are accepted at a 90\%
C.L.

\subsection{Structures in $e^+e^-\to {\rm charmonium}+{\rm hadrons}$}

\begin{figure}[htb]
\centering
     \includegraphics[width=0.32\textwidth,height=0.19\textheight]{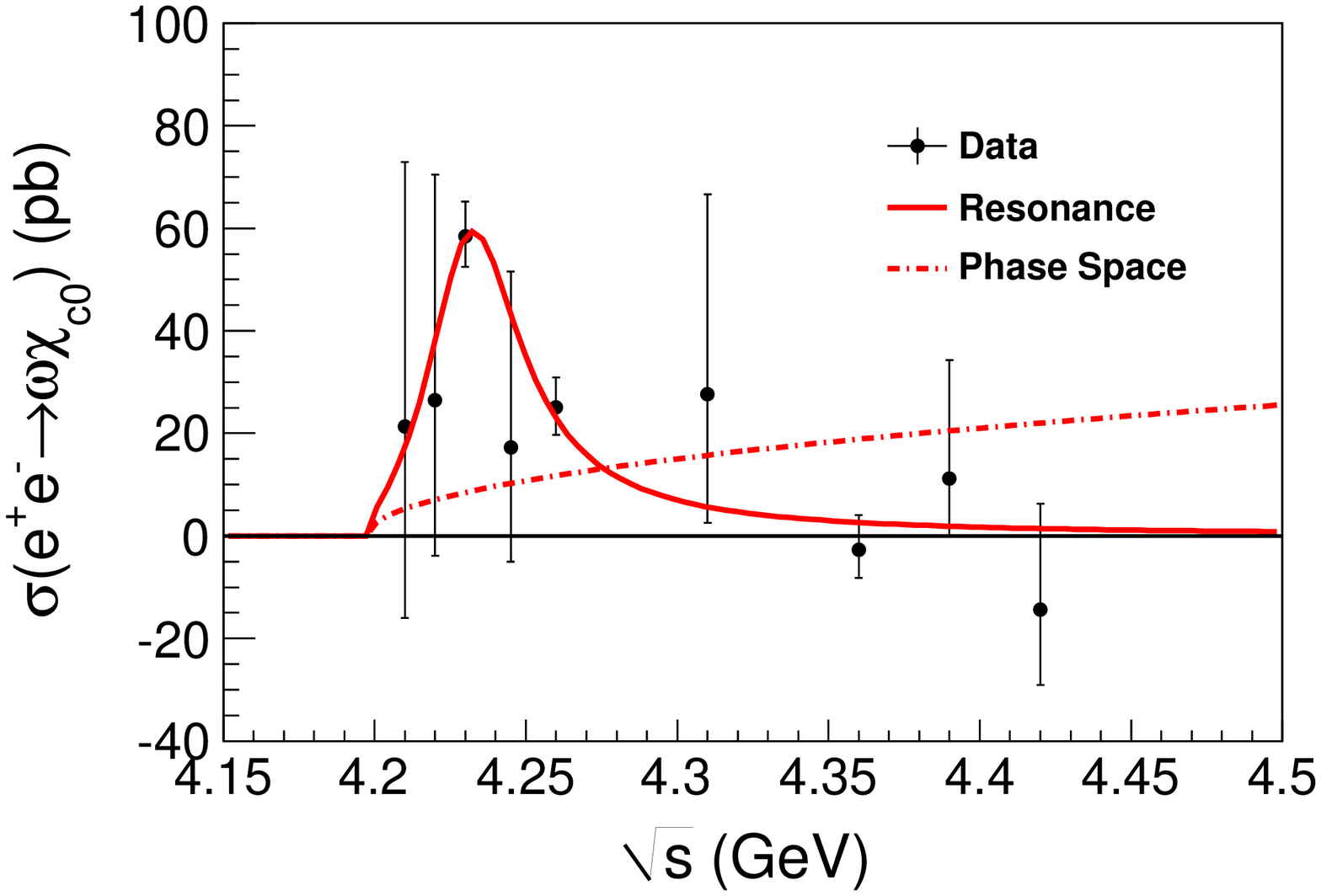}
          \put(-135,6){(a)}
     \includegraphics[width=0.32\textwidth,height=0.19\textheight]{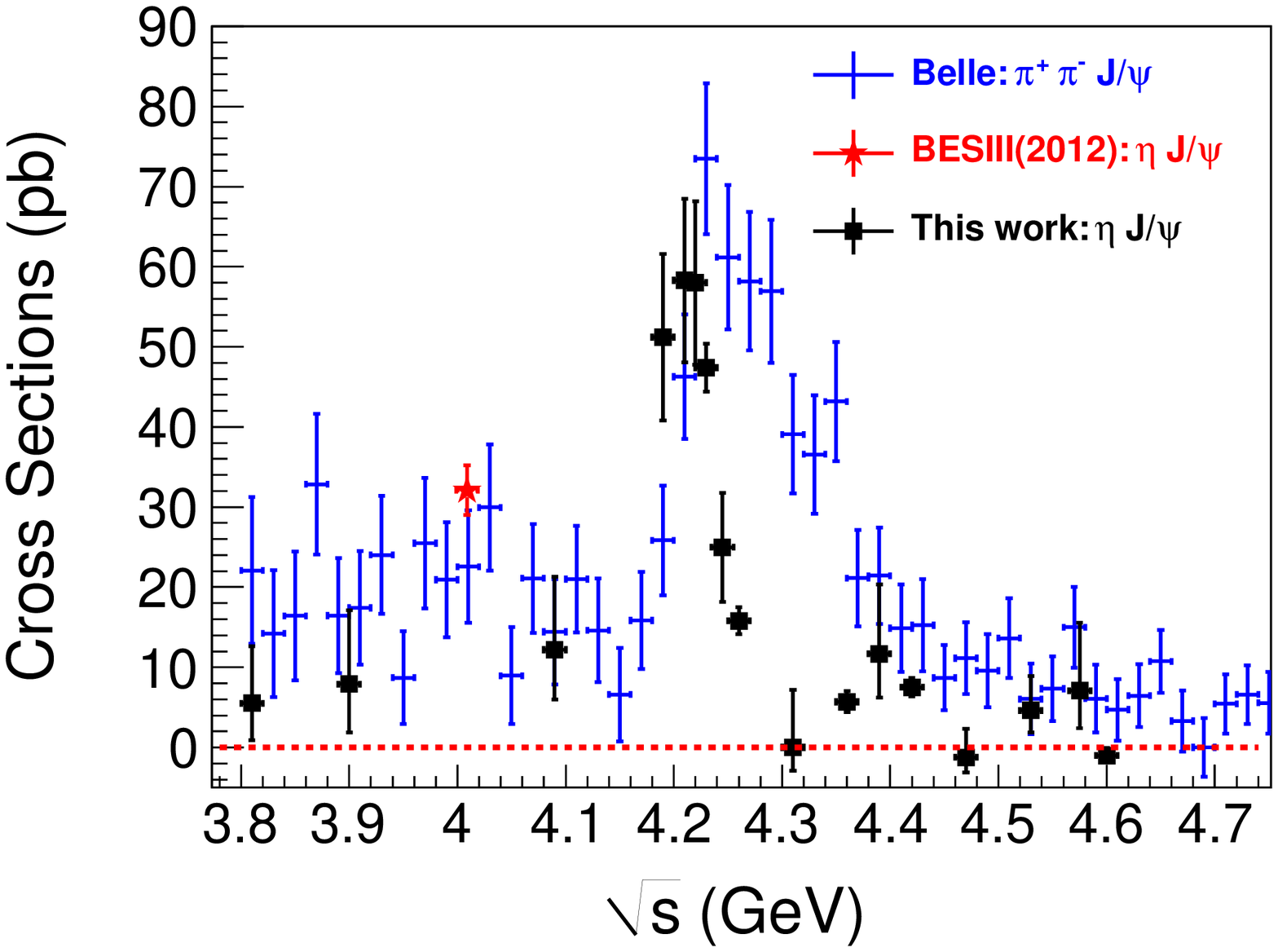}
          \put(-135,6){(b)}
\caption{(a) Fit to $\sigma(e^+e^-\to\omega\chi_{c0})$ with a
resonance (solid curve), or a phase space term (dot-dashed curve).
Dots with error bars are the dressed cross sections. The
uncertainties are statistical only. (b) A comparison of the measured Born cross sections of
$e^{+}e^{-} \to \eta J/\psi$ to those of $e^{+}e^{-}\to
\pi^{+}\pi^{-} J/\psi$ from Belle~\cite{belley_new}. The blue dots are results
of $\pi^{+}\pi^{-} J/\psi$ from Belle.
The errors are statistical only for Belle's results, and are final
combined uncertainties for BESIII's results.} \label{CS-BW-float}
\end{figure}

The  $e^+e^-\to \omega\chi_{c0}$ process was observed at
$\sqrt{s}=4.23$ and 4.26~GeV  for the first time~\cite{Ablikim:2014qwy}. By examining the $\omega\chi_{c0}$ cross section
as a function of center-of-mass energy as shown in Fig.~\ref{CS-BW-float} (a),
we find that it is inconsistent with the line shape of the $Y(4260)$
observed in $e^+ e^-\to\pi^+\pi^-J/\psi$.
Assuming the $\omega\chi_{c0}$ signals
come from a single resonance, we extract
mass and width of the resonance to
be $(4230\pm8\pm6)$~MeV/$c^2$ and $(38\pm12\pm2)$ MeV, respectively,
and the
statistical significance is more than $9\sigma$.

BESIII
analyzed $e^{+}e^{-} \to \eta J/\psi$~\cite{Ablikim:2015xhk}.
Statistically significant $\eta$ signals were observed, and the
corresponding Born cross-sections were measured. The Born cross-sections $\sigma(e^{+}e^{-} \to
\eta J/\psi)$ in this measurement are well consistent with previous
results~\cite{4040,belle}. The measured Born cross-sections
were also compared to those of $e^{+}e^{-} \to \pi^{+}\pi^{-}
J/\psi$ obtained from the Belle experiment~\cite{belley_new}, as
shown in Fig.~\ref{CS-BW-float} (b). Different line shapes can be
observed in these two processes, indicating that the production
mechanism of  $\eta J/\psi$ differs from that of
$\pi^{+}\pi^{-}J/\psi$ in the vicinity of $\sqrt{s}$ =
4.1--4.6~GeV. This could indicate the existence of a rich spectrum
of the $Y$ states in this energy region with different coupling
strengths to various decay modes.

\section{Light meson spectroscopy}
Glueballs and other resonances with large gluonic components are predicted as bound
states by QCD. The lightest (scalar) glueball is estimated to
have a mass in the range from 1 to 2 GeV/$c^2$; pseudoscalar and tensor glueballs are expected
at higher masses. Radiative decays of the charmonium provide a gluon rich
environment and are therefore regarded as one of the most promising hunting grounds for glueballs and hybrids.

\subsection{Observation and Spin-Parity Determination of the $X(1835)$ in $J/\psi\to\gamma K^0_S K^0_S\eta$}
\begin{figure}[htb]
  \centering
     \includegraphics[width=0.32\textwidth,height=0.19\textheight]{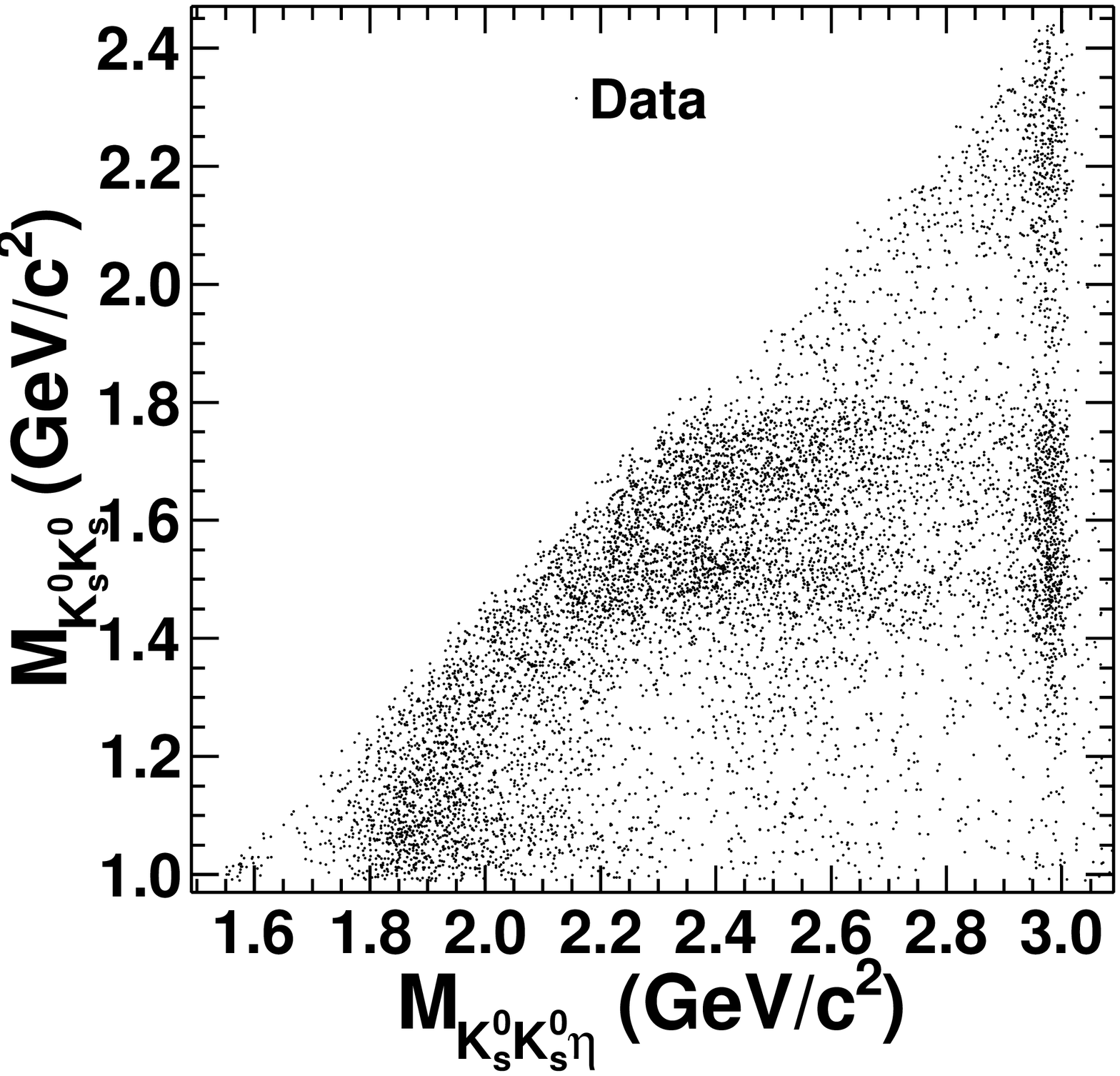}
   \put(-135,6){(a)}
   \includegraphics[width=0.32\textwidth,height=0.19\textheight]{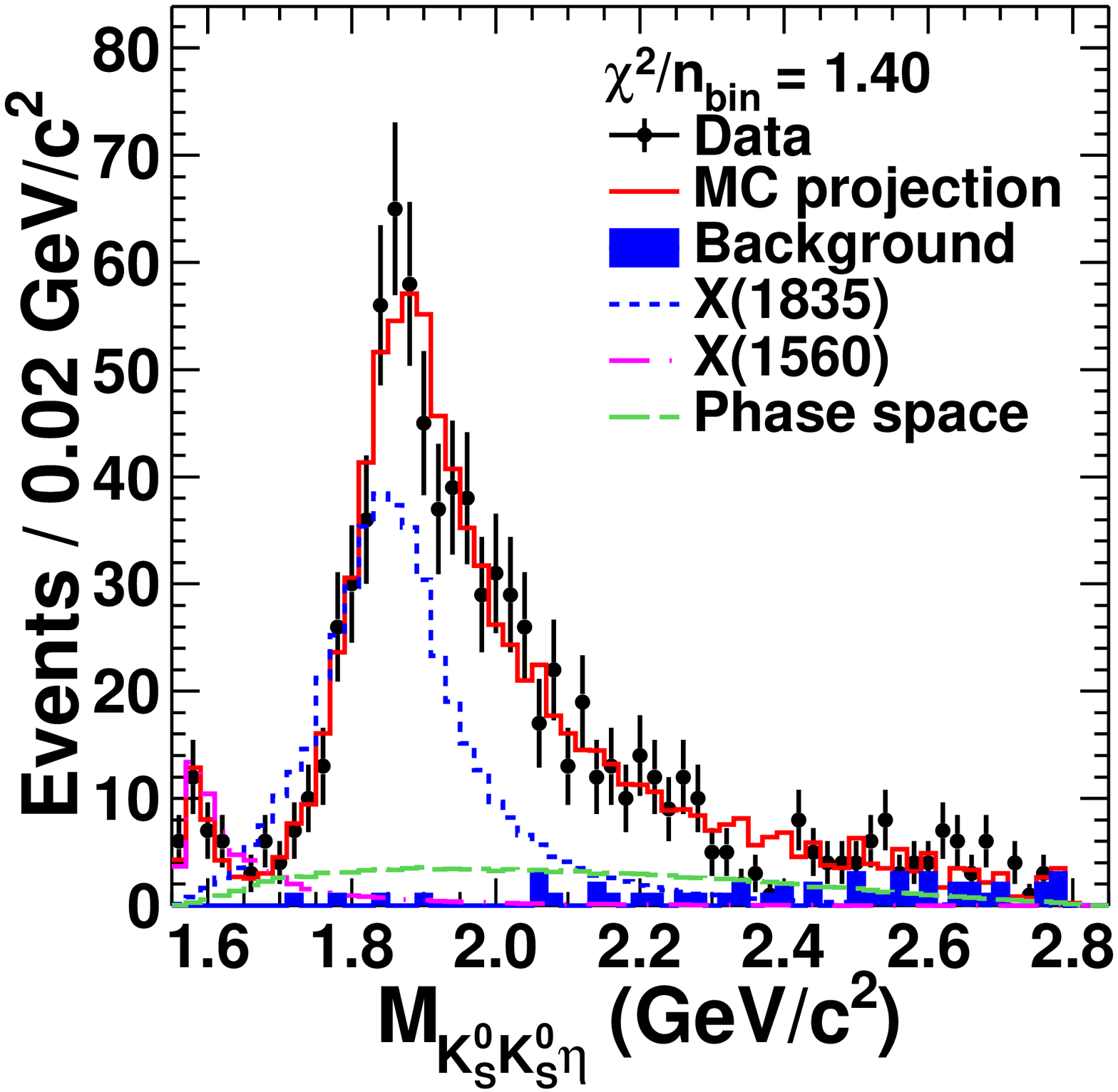}
   \put(-135,6){(b)}
   \includegraphics[width=0.32\textwidth,height=0.19\textheight]{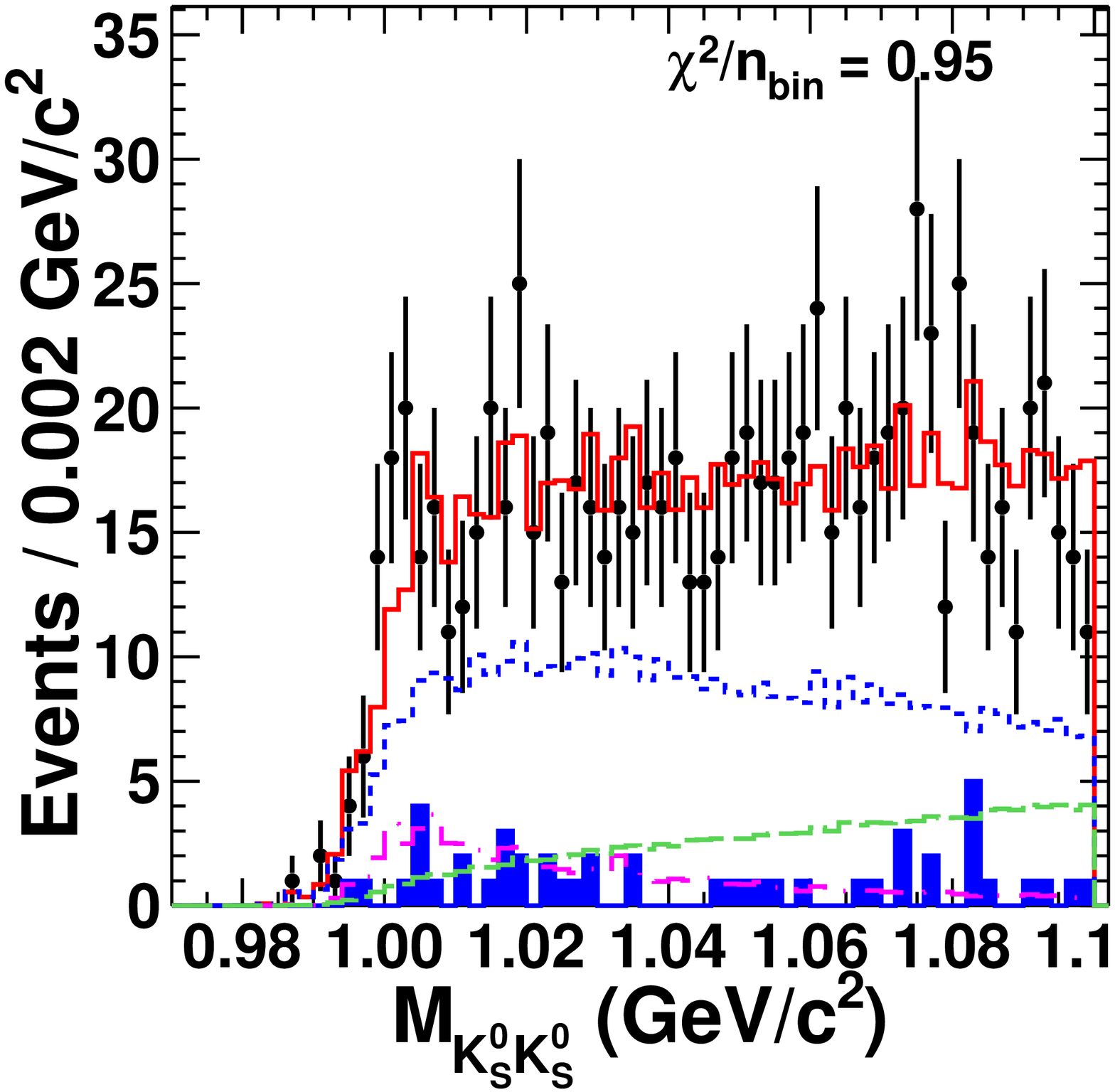}
   \put(-135,6){(c)}
    \caption{\label{KKeta} (a) scatter plot of $M_{ K^0_S K^0_S}$ versus $M_{ K^0_S K^0_S\eta}$;
    Comparisons between data and PWA fit projections. (b), and (c) are the invariant mass distributions of $K^0_S K^0_S\eta$, $K^0_S K^0_S$.
    Dots with error bars
      are data; the shaded histograms are the non-$\eta$ backgrounds
      estimated by the $\eta$ sideband; the solid histograms are phase
      space MC events of $J/\psi\to\gamma K^0_S K^0_S\eta$ with arbitrary
      normalization.}
      \label{fig:gammakketa}
\end{figure}

$X(1835)$ was first observed in $J/\psi\to\gamma \eta^\prime \pi^+ \pi^-$ by
BESII~\cite{x1835_bes2}; this observation was subsequently confirmed by
BESIII~\cite{x1835_bes3}. In addition, an enhancement in
the invariant $p\bar{p}$ mass at threshold, $X(p\bar{p})$, was first
observed by BESII in the decay $J/\psi\to \gamma p \bar{p}$~\cite{gppb_bes2},
and was later also seen by BESIII~\cite{gppb_bes3} and
CLEO~\cite{gppb_cleo}. In a partial wave analysis(PWA), BESIII determined
the $J^{PC}$ of the $X(p\bar{p})$ to be $0^{-+}$~\cite{gppbpwa_bes3}.
The mass of the $X((p\bar{p})$ is consistent with $X(1835)$, but the width of the
$X(p\bar{p})$ is significantly narrower. To understand the nature of the $X(1835)$, it is crucial to measure
its $J^{PC}$ and to search for new decay modes.
Fig.~\ref{gammakketa} (a) shows  the scatter plot of
the invariant mass of ${ K^0_S K^0_S}$ versus that of ${ K^0_S K^0_S\eta}$, indicating the structure around 1.85~GeV/$c^2$ is
strongly correlated with $f_0(980)$.
A partial wave analysis (PWA) of $J/\psi\to\gamma K^0_S K^0_S\eta$ has been performed in the
mass range $M_{K^0_S K^0_S\eta}<2.8$~GeV/$c^2$~after requiring
$M_{K^0_S K^0_S}<1.1$~GeV/$c^2$~\cite{Ablikim:2015toc}. Fig.~\ref{fig:gammakketa} (b) and (c) are the invariant mass distributions of $K^0_S K^0_S\eta$, $K^0_S K^0_S$. Overlaid on the data are the PWA fit
projections, as well as the individual contributions from each
component. The PWA fit requires a contribution from
$X(1835)\to K^0_S K^0_S\eta$ with a statistical significance greater than
12.9$\sigma$, where the $X(1835)\to K^0_S K^0_S\eta$ is dominated by
$f_{0}(980)$ production. The spin parity of the $X(1835)$ is
determined to be $0^{-+}$. The mass and width of the $X(1835)$ are
measured to be $1844\pm 9(stat)^{+16}_{-25}(syst)$ MeV/$c^2$~and
$192^{+20}_{-17}(stat) ^{+62}_{-43}(syst)$ MeV,
respectively. The corresponding product branching fraction
$\mathcal{B}_{X(1835)}$ is measured to be
$(3.31^{+0.33}_{-0.30}(stat)^{+1.96}_{-1.29}(syst)) \times
10^{-5}$.
The mass and width of the $X(1835)$ are consistent with the values
obtained from the decay $J/\psi\to\gamma\eta^\prime\pi^+\pi^-$ by
BESIII~\cite{x1835_bes3}. These results are all first-time
measurements and provide important information to further
understand the nature of the $X(1835)$. Another $0^{-+}$ state, the $X(1560)$, also is observed in data with a
statistical significance larger than 8.9$\sigma$. The mass and width of the $X(1560)$ are consistent with those of the
$\eta(1405)$ and $\eta(1475)$ as given in Ref.~\cite{pdg2014} within
$2.0\sigma$ and $1.4\sigma$, respectively.

\subsection{Amplitude analysis of the  $\pi^{0}\pi^{0}$ system produced in radiative $J/\psi$ decays}
A mass independent amplitude analysis of the $\pi^{0}\pi^{0}$ system in radiative $J/\psi$ decays is
performed~\cite{Ablikim:2015umt}. This analysis uses the world's largest data sample of its type, collected with the BESIII
detector, to extract a piecewise function that describes the scalar and tensor $\pi\pi$ amplitudes in
this decay.  While the analysis strategy employed to obtain results has complications, namely
ambiguous solutions, a large number of parameters, and potential bias in subsequent analyses from
non-Gaussian effects, it minimizes systematic bias arising from
assumptions about $\pi\pi$ dynamics, and, consequently, permits the development of dynamical
models or parameterizations for the data.

The intensities and phase differences for the amplitudes in the fit are
presented as a function of $M_{\pi^{0}\pi^{0}}$ in Ref.~\cite{Ablikim:2015umt}. Additionally, in order to facilitate the development of models, the intensities and phases for each
bin of $M_{\pi^{0}\pi^{0}}$ are given in supplemental materials of Ref.~\cite{Ablikim:2015umt}.
These results may be combined with those of similar reactions for a more comprehensive study of the
light scalar meson spectrum.  Finally, the branching fraction of radiative $J/\psi$ decays to
$\pi^{0}\pi^{0}$ is measured to be $(1.15\pm0.05)\times10^{-3}$, where the error is systematic only
and the statistical error is negligible. This is the first measurement of this branching fraction.


\subsection{Partial Wave Analysis of $J/\psi\to \gamma \phi \phi$}
\begin{figure*}[htb]
   \centering
     \includegraphics[width=0.32\textwidth,height=0.19\textheight]{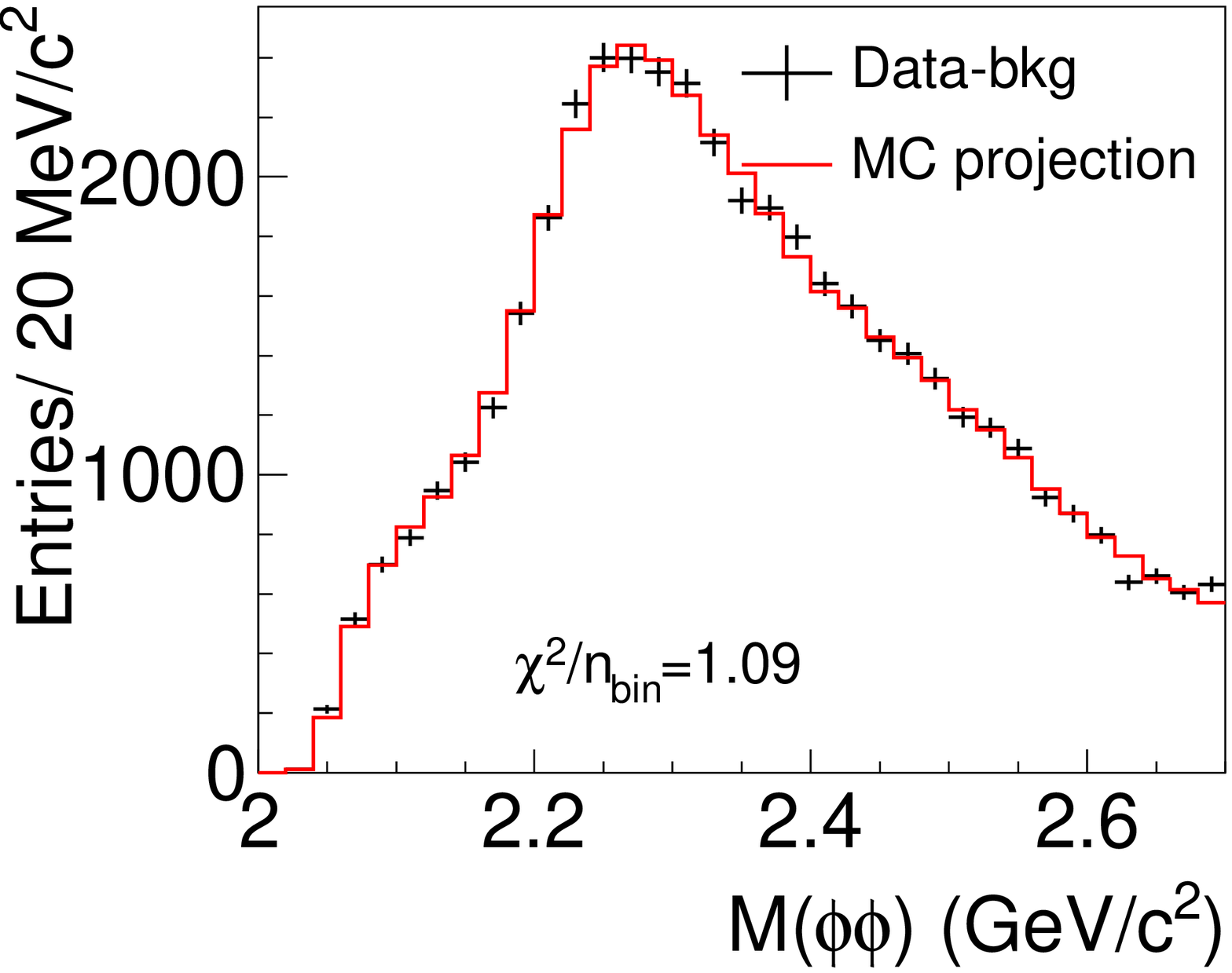}
     \put(-135,6){(a)}
     \includegraphics[width=0.32\textwidth,height=0.19\textheight]{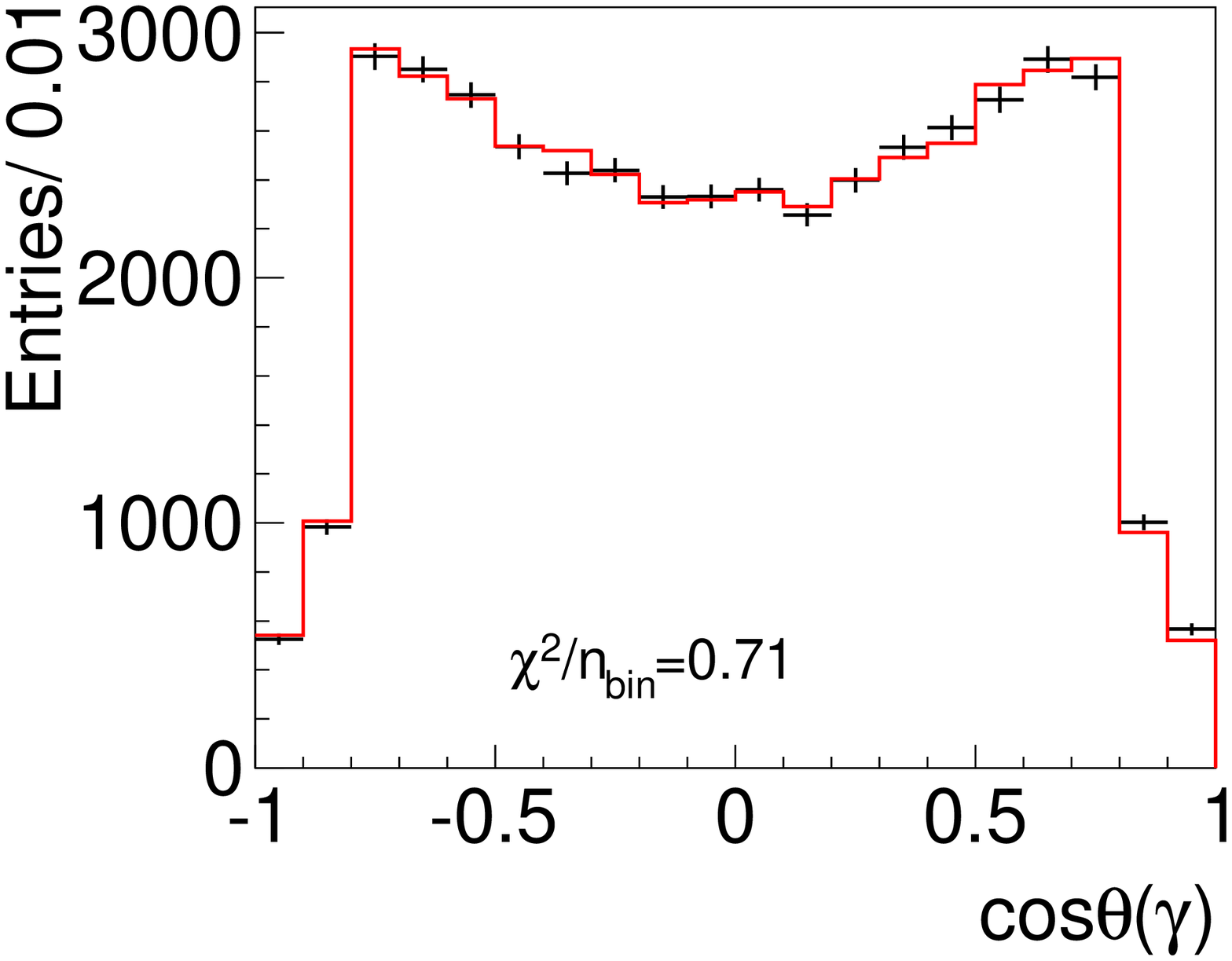}
     \put(-135,6){(b)}
     \includegraphics[width=0.32\textwidth,height=0.19\textheight]{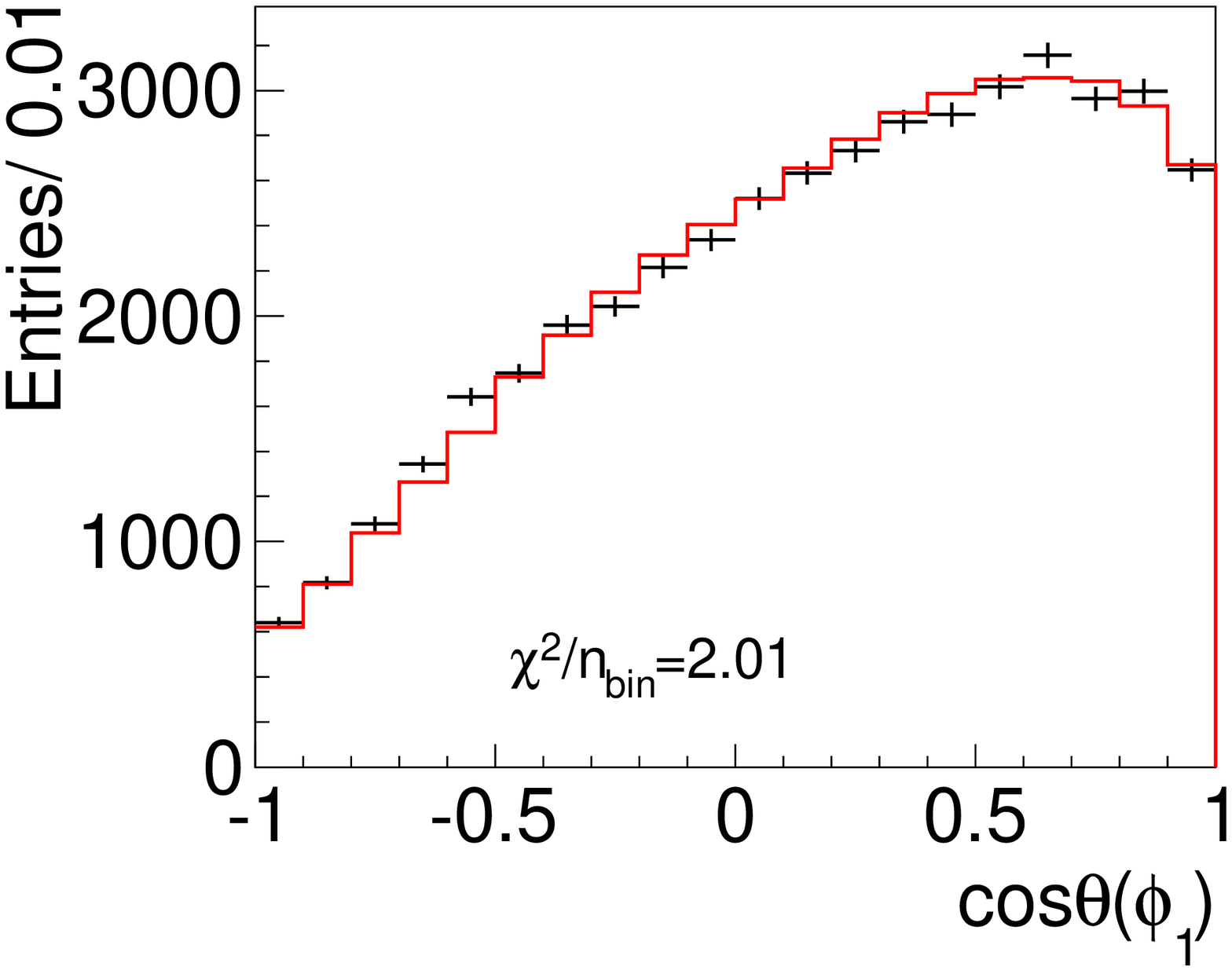}
     \put(-135,6){(c)} \\
    \includegraphics[width=0.32\textwidth,height=0.19\textheight]{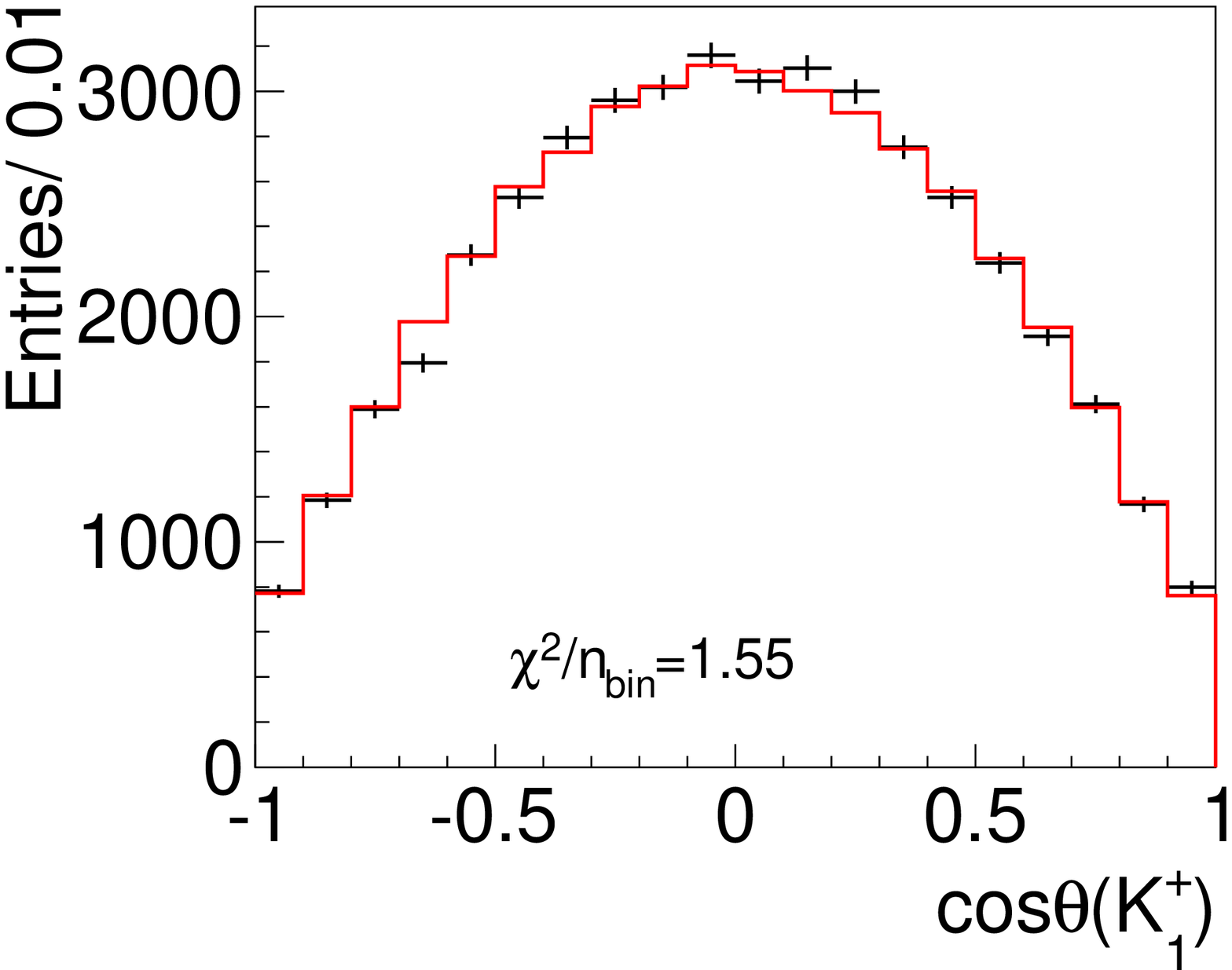}
     \put(-135,6){(d)}
     \includegraphics[width=0.32\textwidth,height=0.19\textheight]{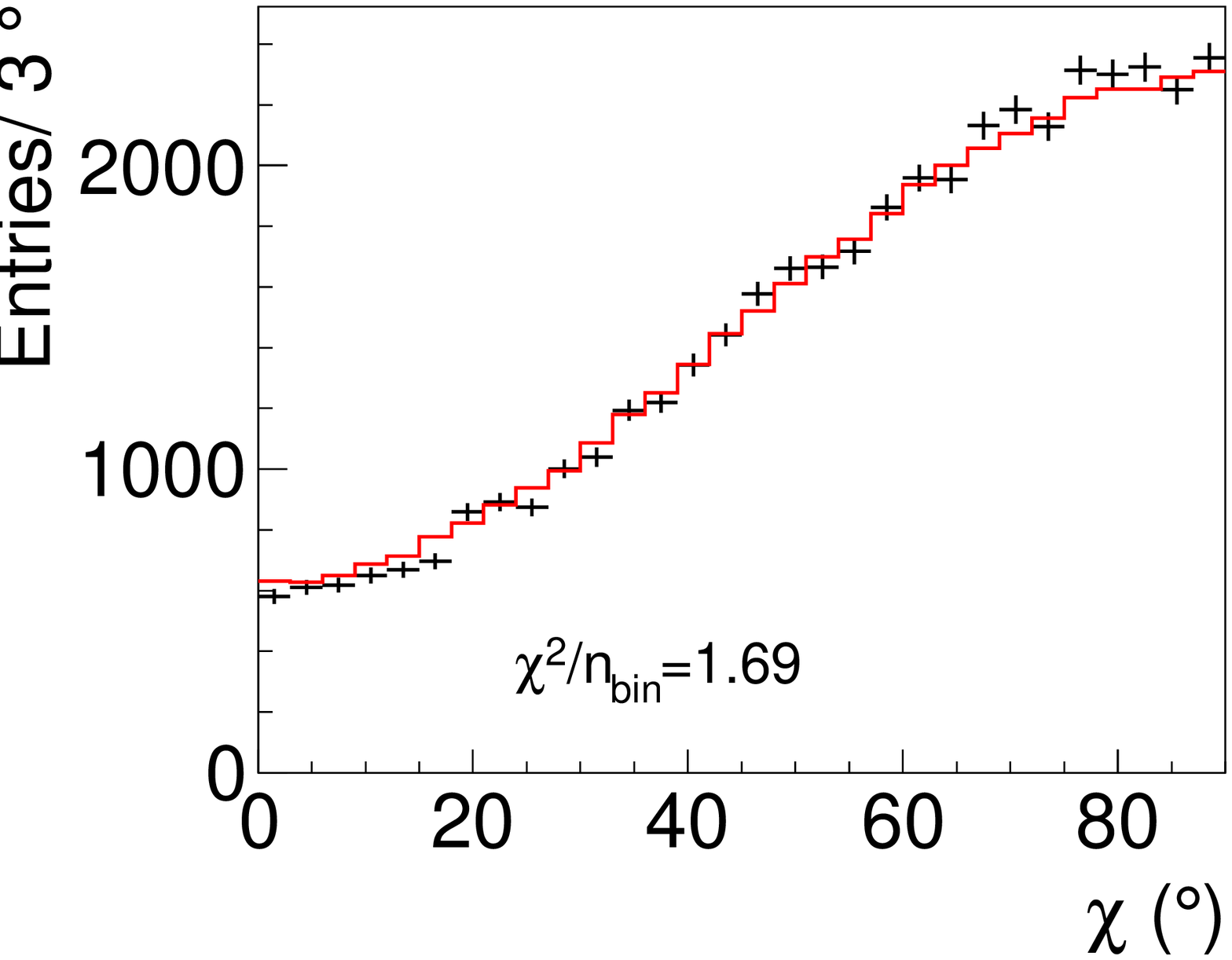}
     \put(-135,6){(e)}
     \includegraphics[width=0.32\textwidth,height=0.19\textheight]{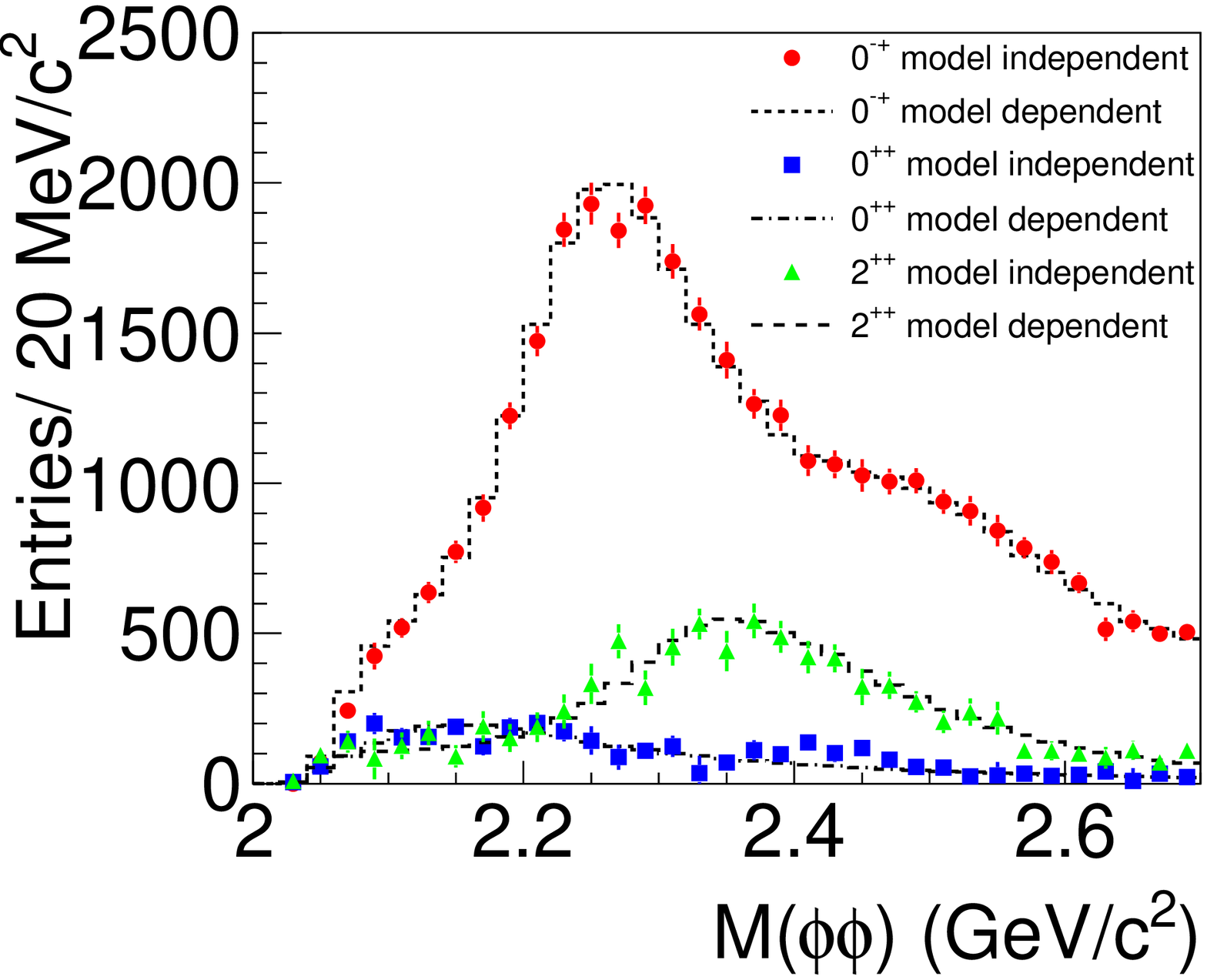}
     \put(-135,6){(f)}
   \caption{Superposition of data and the PWA fit projections for: (a) invariant mass distributions of $\phi\phi$; (b) cos$\theta$ of $\gamma$ in the $J/\psi$ rest frame; (c) cos$\theta$ of $\phi_1$ in the $X$ rest frame; (d) cos$\theta$ of $K^{+}$ in the $\phi_1$ rest frame; (e) the azimuthal angle between the normals to the two decay planes of $\phi$ in the $X$ rest frame. Black dots with error bars are data with background events subtracted and the solid red lines are projections of the model dependent fit. (f) Intensities of individual $J^{PC}$ components. The red dots, blue boxes and green triangles with error bars are the intensities of $J^{PC}$ = $0^{-+}$, $0^{++}$ and $2^{++}$, respectively, from the model independent fit in each bin. The short-dashed, dash-dotted and long-dashed histograms show the coherent superpositions of the BW resonances with $J^{PC}$ = $0^{-+}$, $0^{++}$ and $2^{++}$, respectively, from the model dependent fit.}
   \label{fig:gammaphiphi}
   \end{figure*}

The low lying pseudoscalar glueball is predicted to be around 2.3$-$2.6 GeV/$c^2$ by Lattice QCD ~\cite{bib1,bib2,bib3}.
Aside from the $\eta(2225)$, very little is known in the pseudoscalar sector above 2 GeV/$c^2$. A partial wave analysis of the decay $J/\psi\rightarrow \gamma \phi \phi$ is performed in order to study the intermediate states. The most remarkable feature of the PWA results is
that $0^{-+}$ states are dominant. The existence of the $\eta(2225)$
is confirmed and two additional pseudoscalar states, $\eta(2100)$ with a
mass $2050_{-24}^{+30}$$_{-26}^{+77}$ MeV/$c^{2}$ and
a width $250_{-30}^{+36}$$_{-164}^{+187}$ MeV/$c^{2}$
and X(2500) with a mass $2470_{-19}^{+15}$$_{-23}^{+63}$ MeV/$c^{2}$ and
a width $230_{-35}^{+64}$$_{-33}^{+53}$ MeV/$c^{2}$, are observed.
The new experimental results are helpful for mapping out pseudoscalar
excitations and searching for a $0^{-+}$ glueball.
The three tensors $f_2(2010)$, $f_2(2300)$ and $f_2(2340)$ observed
in $\pi^- p\rightarrow \phi\phi n$~\cite{bibpiN} are also observed
in $J/\psi\rightarrow\gamma\phi\phi$. Recently, the production rate
of the pure gauge tensor glueball in $J/\psi$ radiative decays has been
predicted by Lattice QCD~\cite{bibtensorglueball}, which is compatible
with the large production rate of the $f_2(2340)$
in $J/\psi\rightarrow\gamma\phi\phi$ and $J/\psi\rightarrow\gamma\eta\eta$~\cite{bibgee}. Fig.~\ref{fig:gammaphiphi} shows the PWA fit results with comparison of data.

\section{Light baryon spectroscopy}
The $J/\psi$ and $\psi'$ experiments at BES provide an
excellent place for studying excited nucleons and hyperons -- $N^*$,
$\Lambda^*$, $\Sigma^*$ and $\Xi^*$ resonances\cite{Zou:2000wg}.
Complementary to other facilities, the baryon program at BES3 has several advantages~\cite{Zou:2000wg}. For instance, $\pi N$ and $\pi\pi N$ systems from
$J/\psi\to\bar NN\pi$ and $\bar NN\pi\pi$ processes have an isospin of 1/2 due to isospin conservation; $\psi$ mesons decay to baryon-antibaryon pairs through three or
more gluons, where is a favorable place for producing hybrid (qqqg) baryons,
and for searching some "missing" $N^*$ resonances which have weak
coupling to both $\pi N$ and $\gamma N$, but stronger coupling to $g^3N$. Recently, in a partial wave analysis of $\psi(3686)\to p \bar{p} \pi^0$ \cite{Ablikim:2012zk}, two new $N^*$ resonances $N(2300)$ and $N(2570)$ are observed with $J^P$ assignment of $1/2^+$ and $5/2^-$, respectively.
In the studies of the decays
of $\psi(3686)\rightarrow K^- \Lambda \bar{\Xi}^+ +c.c.$ and $\psi(3686)\rightarrow \gamma K^- \Lambda \bar{\Xi}^+ +c.c.$~\cite{Ablikim:2015apm}, two hyperons, $\Xi(1690)^-$ and
$\Xi(1820)^-$, are observed in the $K^-\Lambda$ invariant mass
distribution in the decay $\psi(3686)\rightarrow K^- \Lambda \bar{\Xi}^+ +c.c.$ with significance of
$4.9 \sigma$ and $6.2 \sigma$, respectively. The results are shown in shown in Fig.~\ref{Xi}. The branching fractions
of $\psi(3686)\rightarrow K^- \Lambda \bar{\Xi}^++c.c.$, $\psi(3686)\rightarrow K^- \Sigma^0 \bar{\Xi}^++c.c.$, $\psi(3686)\to\gamma \chi_{cJ}\to \gamma K^-
\Lambda \bar{\Xi}^+ +c.c.$ $(J=0,\ 1,\ 2)$, and $\psi(3686)\to
\Xi(1690/1820)^{-} \bar{\Xi}^++c.c$ with subsequent decay
$\Xi(1690/1820)^-\to K^-\Lambda$ are measured.


\section{Summary}
With the world¡¯s largest samples of $J/\psi$, $\psi(3686)$, $\psi(3770)$, Y(4260) .etc from $e^+e^-$ production, the BESIII experiment made a significant contribution to the study of the charmonium spectroscopy, light meson spectroscopy and light baryon spectroscopy. BESIII will continue to run 6--8 years. Complementary to other experiments, with various production mechanisms, BESIII will continue shedding light on the the nature of hadrons.



\nocite{*}
\bibliographystyle{aipnum-cp}%
\bibliography{sample}%

\begin{thebibliography}{99}
\bibitem{bes3dect}
M. Ablikim {\it et al.} (BESIII Collaboration), Nucl.\ Instrum.\ Meth.\ A {\bf 614}, 345 (2010).
\bibitem{Fritzsch:1973pi}
  H.~Fritzsch, M.~Jell-Mann and H.~Leutwyler,
  Phys.\ Lett.\ B {\bf 47}, 365 (1973).

\bibitem{Barnes:2005pb}
  T.~Barnes, S.~Godfrey and E.~S.~Swanson,
  Phys.\ Rev.\ D {\bf 72}, 054026 (2005)
\bibitem{Ablikim:2013mio}
  M.~Ablikim {\it et al.}  [BESIII Collaboration],
  Phys.\ Rev.\ Lett.\  {\bf 110}, 252001 (2013)

\bibitem{Liu:2013dau}
  Z.~Q.~Liu {\it et al.}  [Belle Collaboration],
  Phys.\ Rev.\ Lett.\  {\bf 110}, 252002 (2013)

\bibitem{Xiao:2013iha}
  T.~Xiao, S.~Dobbs, A.~Tomaradze and K.~K.~Seth,
  Phys.\ Lett.\ B {\bf 727}, 366 (2013) 
\bibitem{BESIII:2015kha}
 M.~Ablikim {\it et al.} [BESIII Collaboration],
  Phys.\ Rev.\ Lett.\  {\bf 115}, no. 11, 112003 (2015)
  doi:10.1103/PhysRevLett.115.112003
  [arXiv:1506.06018 [hep-ex]].

\bibitem{Ablikim:2013xfr}
  M.~Ablikim {\it et al.}  [BESIII Collaboration],
  Phys.\ Rev.\ Lett.\  {\bf 112}, 022001 (2014)
\bibitem{Ablikim:2015cag}
  M.~Ablikim {\it et al.} [BESIII Collaboration],
  Phys.\ Rev.\ D {\bf 92}, no. 3, 032009 (2015)
\bibitem{Ablikim:2015gda}
  M.~Ablikim {\it et al.} [BESIII Collaboration],
  Phys.\ Rev.\ Lett.\  {\bf 115}, no. 22, 222002 (2015)
  doi:10.1103/PhysRevLett.115.222002
  [arXiv:1509.05620 [hep-ex]].
\bibitem{Ablikim:2013wzq}
  M.~Ablikim {\it et al.} [BESIII Collaboration],
  Phys.\ Rev.\ Lett.\  {\bf 111}, no. 24, 242001 (2013)

\bibitem{Ablikim:2014dxl}
  M.~Ablikim {\it et al.} [BESIII Collaboration],
  Phys.\ Rev.\ Lett.\  {\bf 113}, no. 21, 212002 (2014)
\bibitem{Ablikim:2015swa}
  M.~Ablikim {\it et al.} [BESIII Collaboration],
  arXiv:1509.01398 [hep-ex].

\bibitem{Ablikim:2013emm}
  M.~Ablikim {\it et al.} [BESIII Collaboration],
  Phys.\ Rev.\ Lett.\  {\bf 112}, no. 13, 132001 (2014)

\bibitem{Ablikim:2015vvn}
  M.~Ablikim {\it et al.} [BESIII Collaboration],
  arXiv:1507.02404 [hep-ex].

\bibitem{Ablikim:2013dyn}
  M.~Ablikim {\it et al.} [BESIII Collaboration],
  Phys.\ Rev.\ Lett.\  {\bf 112} (2014) 9,  092001
  doi:10.1103/PhysRevLett.112.092001
  [arXiv:1310.4101 [hep-ex]].
\bibitem{bnote} C.~Z.~Yuan (for the BES and Belle Collaborations),
``Exotic Hadrons,'' arXiv:0910.3138 [hep-ex]. We take 5\% from the
range presented in the paper of $2.3\%<\mathcal{B}[X(3872)\to
\pi^+\pi^- J/\psi]<6.6\%$ at 90\% C.L.

\bibitem{Ablikim:2015dlj}
  M.~Ablikim {\it et al.} [BESIII Collaboration],
  Phys.\ Rev.\ Lett.\  {\bf 115}, no. 1, 011803 (2015)
  doi:10.1103/PhysRevLett.115.011803
  [arXiv:1503.08203 [hep-ex]].
\bibitem{Ablikim:2014qwy}
  M.~Ablikim {\it et al.} [BESIII Collaboration],
  Phys.\ Rev.\ Lett.\  {\bf 114}, no. 9, 092003 (2015)
  doi:10.1103/PhysRevLett.114.092003
  [arXiv:1410.6538 [hep-ex]].

\bibitem{Ablikim:2015xhk}
  M.~Ablikim {\it et al.} [BESIII Collaboration],
  Phys.\ Rev.\ D {\bf 91}, no. 11, 112005 (2015)
  doi:10.1103/PhysRevD.91.112005
  [arXiv:1503.06644 [hep-ex]].

\bibitem{4040}   M.~Ablikim {\it et al.} [BESIII Collaboration],
  ``Observation of $e^{+}e^{-} \to \eta J/\psi$ at
  center-of-mass energy $sqrt{s}=4.009$ GeV,''
  Phys.\ Rev.\ D {\bf 86}, 071101 (2012).
\bibitem{belle}   X.~L.~Wang {\it et al.} [Belle Collaboration],
  ``Observation of $\psi(4040)$ and $\psi(4160)$ decay into $\eta J/\psi$,''
  Phys.\ Rev.\ D {\bf 87}, no. 5, 051101 (2013).
\bibitem{belley_new} Z.~Q.~Liu {\it et al.} [Belle Collaboration],
  ``Study of $e^+e^-\to \pi^+\pi^- J/\psi$ and
  Observation of a Charged Charmoniumlike State at Belle,''
  Phys.\ Rev.\ Lett.\  {\bf 110}, 252002 (2013).


\bibitem{Brambilla:2014jmp}
  N.~Brambilla {\it et al.},
  Eur.\ Phys.\ J.\ C {\bf 74}, no. 10, 2981 (2014)
\bibitem{x1835_bes2} M.~Ablikim {\it et al.} (BES Collaboration), Phys. Rev. Lett. {\bf95}, 262001 (2005).
\bibitem{x1835_bes3} M.~Ablikim {\it et al.} (BESIII Collaboration), Phys. Rev. Lett. {\bf106}, 072002 (2011).
\bibitem{gppb_bes2} J.~Z.~Bai {\it et al.} (BES Collaboration), Phys. Rev. Lett. {\bf91}, 022001 (2003).
\bibitem{gppb_bes3} M.~Ablikim {\it et al.} (BESIII Collaboration), Chin. Phys. C {\bf34}, 421 (2010).
\bibitem{gppb_cleo} J.~P.~Alexander {\it et al.} (CLEO Collaboration), Phys. Rev. D {\bf82}, 092002 (2010).
\bibitem{gppbpwa_bes3} M.~Ablikim {\it et al.} (BESIII Collaboration), Phys. Rev. Lett. {\bf108}, 112003 (2012).
\bibitem{Ablikim:2015toc}
  M.~Ablikim {\it et al.} [BESIII Collaboration],
  Phys.\ Rev.\ Lett.\  {\bf 115} (2015) 9,  091803
  doi:10.1103/PhysRevLett.115.091803
  [arXiv:1506.04807 [hep-ex]].
\bibitem{pdg2014} K.~A.~Olive {\it et al.} (Particle Data Group), Chin. Phys. C {\bf38}, 090001 (2014).
\bibitem{Ablikim:2015umt}
  M.~Ablikim {\it et al.} [BESIII Collaboration],
  Phys.\ Rev.\ D {\bf 92}, no. 5, 052003 (2015)
  doi:10.1103/PhysRevD.92.052003
  [arXiv:1506.00546 [hep-ex]].
\bibitem{bib1} G. S. Bali, {\it et al}. (UKQCD Collaboration), Phys. Lett. B {\bf 309}, 378 (1993).
\bibitem{bib2} C. J. Morningstar and M. Peardon, Phys. Rev. D {\bf 60}, 034509 (1999).
\bibitem{bib3} Y. Chen {\it et al}., Phys. Rev. D {\bf 73}, 014516 (2006).
\bibitem{bibpiN} A. Etkin {\it et al}., Phys. Rev. Lett. {\bf 41}, 784 (1978); Phys. Lett. B {\bf 165}, 217 (1985); Phys. Lett. B {\bf 201}, 568 (1988).
\bibitem{bibtensorglueball} Y. Chen {\it et al}., Phys. Rev. Lett. {\bf 111}, 091601 (2013).
\bibitem{bibgee} M. Ablikim {\it et al}. (BESIII Collaboration), Phys. Rev. D {\bf 87}, 092009 (2013).
\bibitem{Zou:2000wg}
  B.~S.~Zou,
  Nucl.\ Phys.\ A {\bf 684}, 330 (2001)
  doi:10.1016/S0375-9474(01)00433-X
  [hep-ph/0006039].
\bibitem{Ablikim:2012zk}
  M.~Ablikim {\it et al.} [BESIII Collaboration],
  Phys.\ Rev.\ Lett.\  {\bf 110}, no. 2, 022001 (2013)
  doi:10.1103/PhysRevLett.110.022001
  [arXiv:1207.0223 [hep-ex]].
\bibitem{Ablikim:2015apm}
  M.~Ablikim {\it et al.} [BESIII Collaboration],
  Phys.\ Rev.\ D {\bf 91}, no. 9, 092006 (2015)
  doi:10.1103/PhysRevD.91.092006
  [arXiv:1504.02025 [hep-ex]].
\end{thebibliography}

\end{document}